\documentclass[%
 reprint,
 amsmath,amssymb,
 aps, 12pt, one column,
superscriptaddress,showkeys,
noeprint 
]{revtex4-2}

\usepackage{graphicx}
\usepackage{dcolumn}
\usepackage{bm}

\usepackage{xcolor}
\usepackage{gensymb}
\usepackage{color}
\usepackage{tikz}
\usetikzlibrary{shapes}
\usepackage[english]{babel}

\usepackage{graphicx} 
\usepackage{subcaption}
\usepackage{comment}
\newcommand{\MDG}[1]{\textcolor{magenta}{ *** #1 ***}}

\begin{document}

\preprint{APS/123-QED}

\title{Effect of polymer additives on dynamics of water level in an open channel}


\author{Manish Kumar}
\affiliation{Department of Chemical and Biological Engineering, University of Wisconsin-Madison, 1415 Engineering Dr, Madison, WI 53706, USA}
\author{Michael D. Graham}
\affiliation{Department of Chemical and Biological Engineering, University of Wisconsin-Madison, 1415 Engineering Dr, Madison, WI 53706, USA}

\begin{abstract}

The presence of a tiny amount of polymers (a few parts per million) in a fluid dramatically reduces turbulent drag. 
For this reason, polymer additives have been proposed to be used in flood remediation: 
in an open channel at fixed flow rate, the decrease in friction due to polymer addition is expected to lead to a decrease in water height in the channel. 
However, in a recent field experiment, a counterintuitive transient \textit{increase} in water height has been observed far downstream of polymer injection.
We numerically investigate the effect of polymer additives on the water height in a long canal using the shallow water equations augmented with an evolution equation for polymer concentration that incorporates turbulent dispersion and polymer degradation. Just downstream of polymer injection, the water height decreases due to the decreased friction at fixed volumetric flow rate. Further downstream, however, the height can increase, consistent with the experimental observation. We elucidate the mechanism of this unexpected rise in water height in the canal. We also suggest a technique to mitigate the water rise, as this is detrimental for practical applications. \\

\end{abstract}

\maketitle

\section{Introduction}
The drainage capacity is a core feature of open channels such as irrigation canals and sewer systems. The cross-section area and the slope of the channel's bed are the main design parameters that control the drainage capacity. The slope of a channel is often fixed by the geographical topology and a channel with a larger cross-section is economically expensive to build. Moreover, these control parameters can not be changed once the channel is built, resulting in limited control over the drainage capacity. In fact, for natural channels like rivers and streams, the drainage capacity is naturally fixed which leads to flooding in the case of excess water supply due to heavy rain. 

The drainage capacity of a channel can be improved by the frictional drag reduction of the flow \cite{Abubakar2014}. The addition of a tiny amount ($\sim 20 $ ppm) of high molecular weight polymers such as polymethylmethacrylate (PMMA) significantly reduces turbulent drag \cite{toms1949some}. Therefore, it is widely used in pipeline transport of liquids such as crude oil, water heating and cooling systems, and airplane tank filling \cite{Brostow2008}. Significant efforts have been made to understand the character and mechanism of turbulent drag reduction in pipe flows due to polymer additives \cite{graham2004drag,Graham:2014uj,White2008,Xi2019,Dubief2023}. The stretching of polymeric chains in turbulent flows weakens near wall vortices \cite{Li:2007ii,Kim:2007dq}, which reduces turbulent Reynolds stresses and modifies the mean velocity profile, leading to a global drag reduction of the flow. The turbulent drag initially decreases with polymeric concentration, however, it saturates for a concentration larger than a critical value ($\sim O(10)$ ppm), known as maximum drag reduction (MDR) asymptote \cite{Virk1967,Virk1971}. The drag at MDR asymptote for pipe flows can be achieved as low as $20 \%$ of the Newtonian turbulent drag \cite{herzhaft2000additifs,White2018}. However, it is always greater than the Newtonian laminar drag, because the flow does not completely laminarize and certain forms of turbulence still exist at MDR asymptote \cite{Hof.2013,Xi2019,Shekar:2019hq}.    

Despite the important applications in irrigation, sewer systems, and flood water disposal, the flow dynamics of polymer additives in open channels have received relatively little attention compared to pressure-driven flows. Laboratory scale experiments have been performed to investigate the effect of wall roughness, obstacles, and turbulence type on the turbulent drag reduction due to polymer additives in open channel flows \cite{petrie2003polymer,Janosi2004,Mignot2019}. The flow of dilute polymeric solution through open smooth and rough channels reveals that the polymer addition is more efficient for drag reduction in a smooth channel compared to a rough one \cite{petrie2003polymer,Mignot2019}. In the presence of large obstacles, the drag reduction due to polymer additives is negligible \cite{Mignot2019} because the flow resistance is dominated by form drag due to obstacles and the friction drag due to the boundary layer is negligible \cite{Cadot1998}. Experiments on dam-break flows show the effect of polymer addition on transient current \cite{Janosi2004}. In these experiments, the fluid after the dam break was allowed to flow through channels having a dry bed and a preexisting fluid layer, where the turbulent drag reduction was observed only for the dry bed channel \cite{Janosi2004}. {The flow over a preexisting fluid layer after the dam break generates unstable jets, bubbles, and breaking waves, which are the characteristics of multi-phase turbulence. But, polymeric-induced turbulent drag reduction is more efficient in single-phase turbulence than multi-phase turbulence due to a smooth boundary layer excitation in the single-phase turbulence \cite{Cadot1998}.} 
Therefore, the polymer additives in dam-break flows are more effective for the flow over a dry bed after the dam break compared to the flow over a preexisting fluid layer \cite{Janosi2004}.

A few field experiments have been performed to explore the turbulent drag reduction in open channel flows \cite{sellin1988application,Hart2012,Bouchenafa2021}. The addition of polymer in wastewater enhances the flow speed and reduces the height of the water level leading to the enhancement of the capacity of the sewer systems. During the 2010 Winter Olympics, polymer additives were used in the sewer systems of the main event venue, the resort municipality of Whistler, to enhance their capacity \cite{Hart2012}.  Recently, a field experiment has been reported in a long irrigation canal ($26.3$ km) exploring the effect of polymer additives on the dynamics of water level \cite{Bouchenafa2021}. This experiment lasted around $18$ hours, and $10.5$ tons of polyacrylamide (PAM) was injected into the canal over $15$ hours. The discharge rate through the canal was fixed at $\rm{Q=9.653 \ m^3/s}$ and the polymer additives were added at a rate of $\rm{0.194 \ kg/s}$, resulting in a negligible change in the fluid density. The water height just downstream of the injection point in the canal decreased due to drag reduction and fixed volumetric flow rate. However, further downstream of the canal, the water height increased and remained higher than the base height for several hours, which is a counter-intuitive result and detrimental to applications such as flood remediation. In the present work, we numerically explore the dynamics of the water level in the canal and provide an explanation for the rise of the water level far downstream of the injection point in the canal. We have also suggested a technique to mitigate the rise of water level downstream of the injection point.

\section{Model formulation}

We will consider a one-dimensional (1D) model of the flow where the dynamics of water level are given by the Saint-Venant (shallow-water) equations \cite{de1871theorie}. These represent  conservation of mass and momentum  and can be given as: 
\begin{equation}\label{com}
\frac{\partial h} {\partial t}+\frac{\partial hu} {\partial x}=0,
\end{equation}
\begin{equation}\label{colm}
\frac{\partial u} {\partial t}+u\frac{\partial u} {\partial x}+\frac{1} { Fr^2} \frac{\partial h} {\partial x}+\frac{1} { Fr^2} \left(S_f-S\right)=0,
\end{equation}
 where $h$ and $u$ are the nondimensional water height and velocity, respectively. The water height ($H$) and velocity ($U$) before the injection of the polymer additives in the channel have been used as characteristic height and characteristic velocity to normalize the respective variables throughout the study. The time has been normalized with characteristic time scale $H/U$. The dimensionless canal length is given by $l=L/H$. The Froude number $Fr=U/\sqrt{gH}$ where $g$ is the gravitational acceleration, is a dimensionless number that represents the ratio of inertial to gravitational forces. The bed slope and the frictional slope of the channel have been denoted by $S$ and $S_f$, respectively. The relation between frictional slope and water velocity can be given by Manning's equation \cite{yen1992dimensionally}:
 \begin{equation}\label{s_f}
S_f=\frac{n^2 u^2} {R^{4/3}}f_d,
\end{equation}
where $n$ and $R$ are the Manning coefficient and the normalized hydraulic radius of the channel, respectively. The hydraulic radius of a channel is the ratio of the cross-section area of the flow ($A$) to its wetted perimeter ($P$). For the trapezoidal channel with a 1:1 bank slope, which has been considered in the present study, the hydraulic radius can be given as:  
 \begin{equation}\label{R_hydro}
R=A/P=\frac{h(1+h/w)}{1+2\sqrt{2}h/w},
\end{equation}
where $w=W/H$ is the dimensionless bottom width of the channel. {The values of characteristic height ($H$) and velocity ($U$) have been obtained by solving dimensional governing equations for fully developed flow (i.e., $S_f-S=0$), where volumetric flow rate ($Q$), bed slope ($S$), bottom width ($W$), and Manning coefficient ($n$) determine the base water height and velocity in the channel (Table. \ref{parameters_range}). The experimental values of these parameters lead to  $H=1.02$ m and $U=0.52$ m/s as characteristic scales, which are well within the range of water height and velocity measured in the experiment (Table \ref{parameters_range}). 
} 

\begin{figure}[!ht]
\centering
\begin{subfigure}[b]{0.48\textwidth} 
\includegraphics[width=\textwidth]{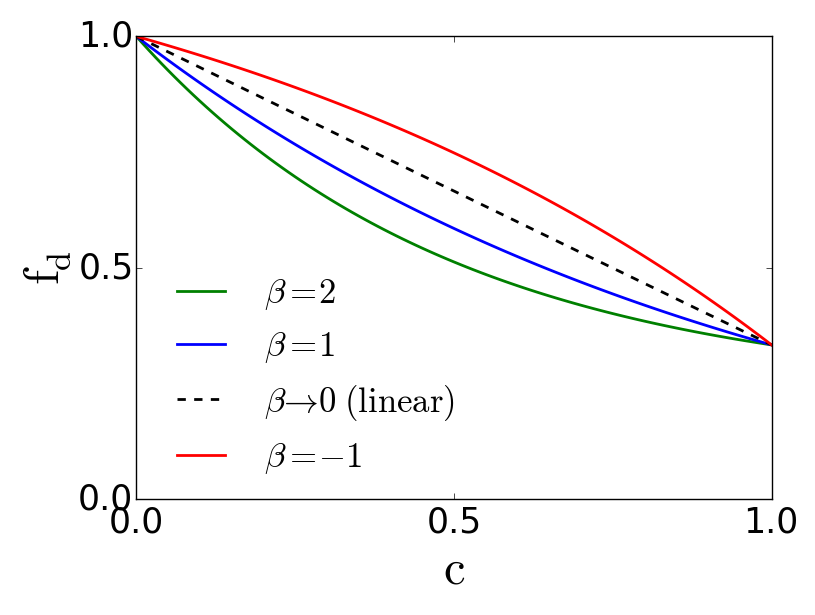}
\caption{}
\label{f_d_vs_c.png}
\end{subfigure}
\begin{subfigure}[b]{.48\textwidth}
\includegraphics[width=\textwidth]{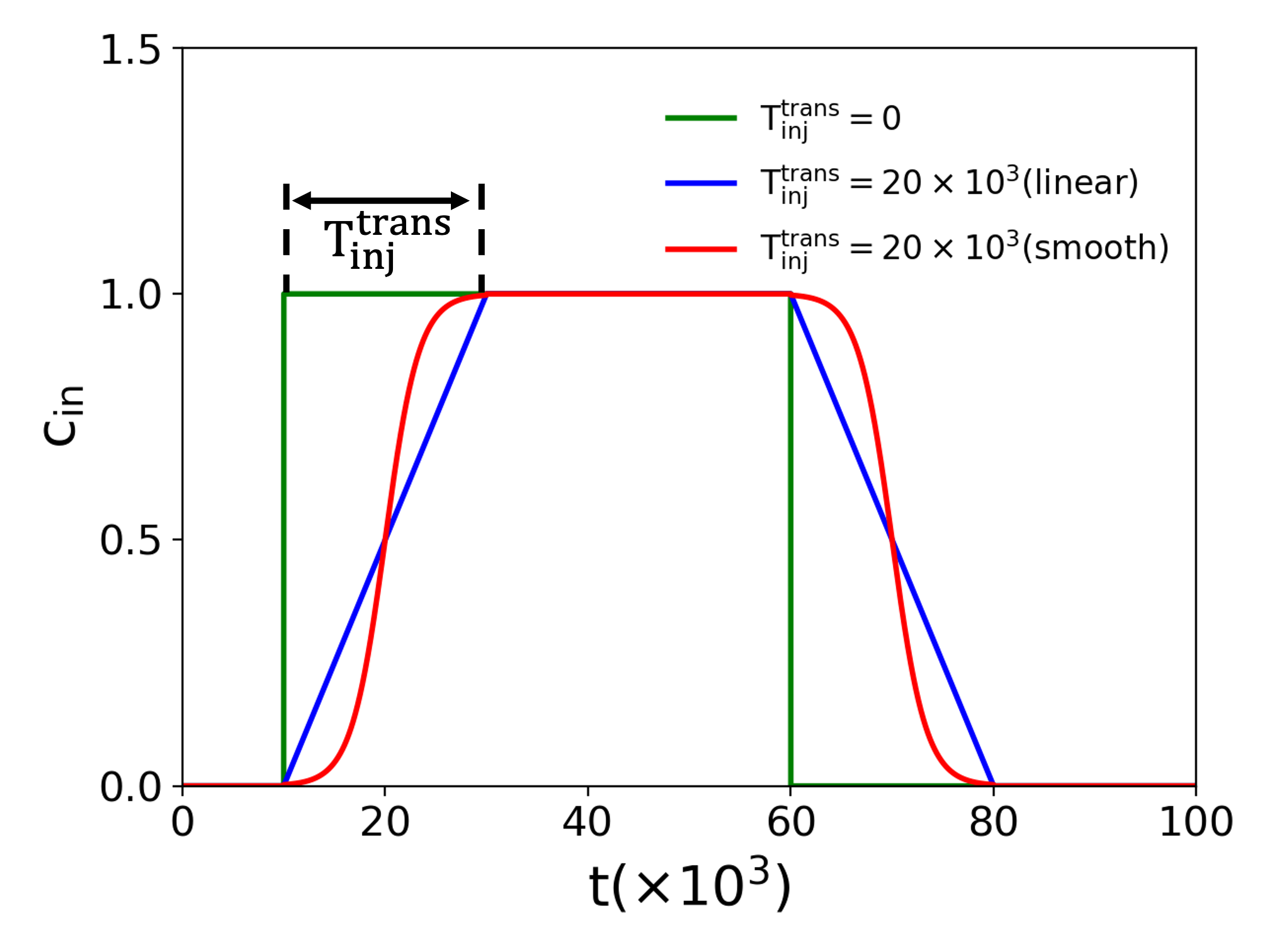}
\caption{}
\label{cin_vs_t_Tinj20k_tnond.png}
\end{subfigure}
\caption{(a) Different profiles of concentration dependent turbulent drag function ($f_d$). (b) a cartoon showing sudden and gradual changes in injection polymeric concentration.}
\end{figure}

The function $f_d$ (Eq. \ref{s_f}) is a concentration-dependent function that represents the relative drag in the presence of polymer additives, where $f_d=1$ in the absence of polymer. For polymer additives, $f_d$ can be described with an exponential profile as \cite{Choueiri2018}:
\begin{equation}\label{fd_vs_c_exp}
f_d=1-\alpha (1-e^{-\beta c}),
\end{equation}
where $\alpha=DR_{max}/(1-\exp(-\beta))$. {Here, $c$ and $DR_{max}$ represent normalized polymeric concentration and the maximum drag reduction in the channel, respectively.} 
The polymeric concentration at the injection point in the experiments of \cite{Bouchenafa2021} ($C_0= 20$ ppm) is sufficient to achieve the maximum drag reduction limit and it has been used to normalize the concentration. The positive and negative values of $\beta$ lead to upward concave and convex profiles of $f_d$, respectively (Fig. \ref{f_d_vs_c.png}). Unless specified otherwise, we take $f_d$ to decrease linearly (i.e., $\beta \to 0$) as polymer concentration increases.  The evolution of polymeric concentration is given by the depth-averaged concentration transport equation as:
\begin{equation}\label{coc}
\frac{\partial c} {\partial t}+u\frac{\partial c} {\partial x}-\frac{1} {Pe} \frac{1} {h} \frac{\partial } {\partial x} \left( h\frac{\partial c} {\partial x}\right)+B c=0,
\end{equation}
where the P\'eclet number $Pe=HU/D_L$ represents the ratio of advective to diffusive transport. The longitudinal turbulent dispersion coefficient has been denoted by $D_L$ and we have used the correlation developed in Ref. \cite{Kashefipour2002} to estimate the range of longitudinal dispersion relevant to the present study. 
The P\'eclet number is based on the water height ($H$) in the channel and the channel length is much larger than the height ($l\sim 10^4$). Therefore, although $Pe<1$, we do not expect any significant influence of turbulent dispersion on the dynamics of the water level in the channel because $Pe l\gg 1$. The turbulent nature of the flow induces the degradation of the polymeric chains \cite{Moussa1994,Pereira2012}. Here the ratio of the polymeric degradation rate to the shear rate has been denoted by $B$. The polymeric degradation rate is not known in the experiment. However, we note that the effect of polymer additives on the steady-state height of the water level vanishes at the exit of the channel for $B>3 \times 10^{-4}$ (Appendix \ref{tool_valid}). Therefore, we use the range $B=0-4\times 10^{-4}$ to explore the effect of polymer degradation on the dynamics of water level in the present study. 

The mass and momentum conservation equations (Eqs. \ref{com} and \ref{colm}) contain the first-order derivatives of $u$ and $h$ requiring a single boundary condition for each. The volumetric flow rate through the channel is constant leading to the following boundary condition at the inlet:
 \begin{equation}\label{u_bc_inlet}
q=uh(w+h),
\end{equation}
where $q$ is the normalized volumetric flow rate through the channel. In the present study, the flow is subcritical as $Fr<1$ (Table \ref{parameters_range}). Therefore, the boundary condition for $h$ needs to be prescribed at the exit of the channel ($x=l$) \cite{CHANSON2004302,Soulis1991,Delestre2013}. We consider a zero-gradient boundary condition for $h$ at the exit as the channel is very long. The boundary conditions for the concentration equation (Eq. \ref{coc}) are given as:
 \begin{equation}\label{c_bcs}
c=c_{in} \ {\rm and} \ \frac{\partial c}{\partial x}=0
\end{equation}
at the channel inlet ($x=0$) and exit ($x=l$), respectively. 

The time-dependent profiles of injection concentration $c_{in}$ used in the present study are shown in Fig. \ref{cin_vs_t_Tinj20k_tnond.png}. Unless specified otherwise, we use the injection profile having a sudden change in the concentration ($T_{inj}^{trans}=0$) similar to the experiment \cite{Bouchenafa2021}. The other profiles shown are implemented to determine whether they can mitigate the observed downstream height increase. For the numerical simulation, the governing equations have been discretized using the finite difference method. An upwind scheme has been used for the discretization of advective terms and the Euler method has been used for the time stepping. The validation of the numerical tool and the mesh-dependence study have been given in Appendix \ref{tool_valid}. 

\begin{table}[h!]
 \caption{ \label{parameters_range} The values of different parameters associated with the experimental data \cite{Bouchenafa2021}.}
 \begin{center}
 \begin{tabular}{ |c|c|c|c|c|c|c|c|c| } 
 \hline
 $Fr$ & $Pe$ & $l (\times 10^3)$ & $S$ & $n$ & $Q (m^3/s)$ &$U (m/s)$ & $H (m)$ & $W (m)$\\
 \hline
 $0.1-0.2$ & $0.002-0.02$ & $17.5-31$ & $10^{-4}$ & $0.0094$ & $9.653$ & $0.35-0.64$ & $0.85-1.5$& $17$ \\
 \hline
 
 \end{tabular}
\end{center}
\end{table}

\section{Results and discussion}
\subsection{Linear dynamics}

 \begin{figure}[!ht]
\centering
\begin{subfigure}[b]{0.48\textwidth} 
\includegraphics[width=\textwidth]{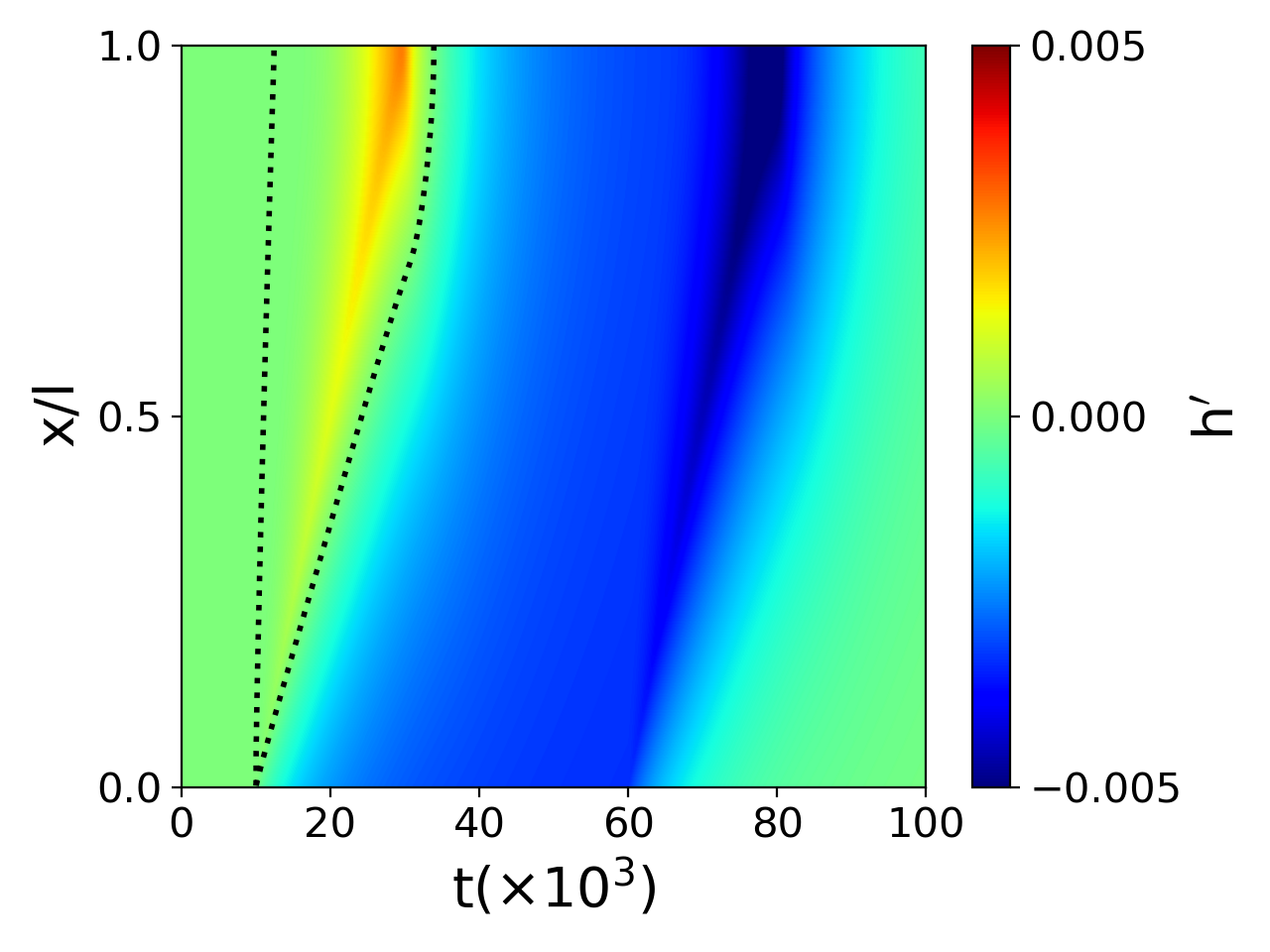}
\caption{}
\label{hprime_vs_x_and_t}
\end{subfigure}
\begin{subfigure}[b]{.48\textwidth}
\includegraphics[width=\textwidth]{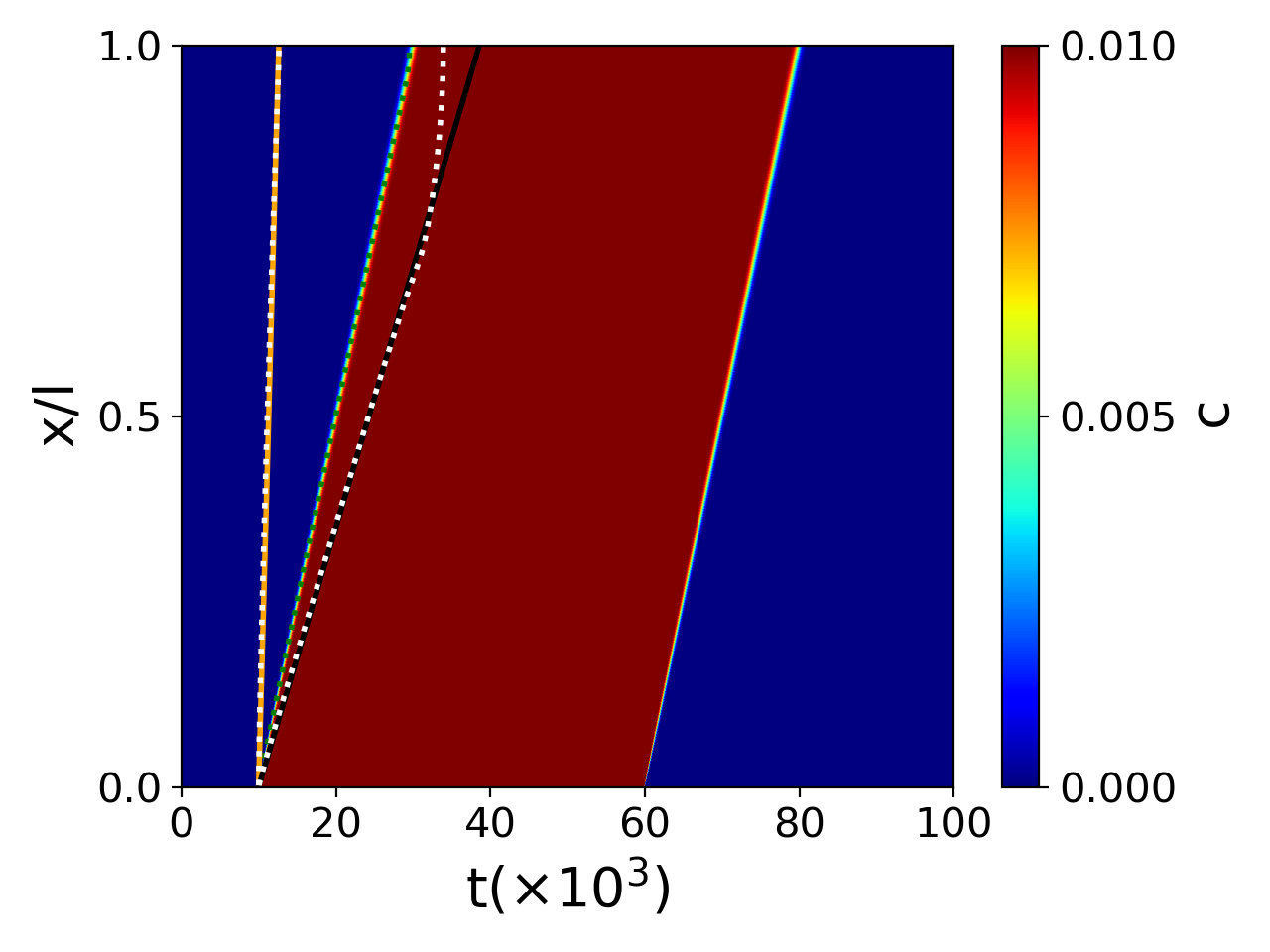}
\caption{}
\label{cprime_vs_x_and_t}
\end{subfigure}
\caption{Spatiotemporal contours of (a) height perturbation and (b) polymeric concentration for a tiny injection concentration of polymer additives ($c_{in}=0.01$). The space enclosed by black dotted lines in (a) and white dotted lines in (b) represents the region where the water level is higher than the base height during the polymer injection. The solid orange line (b) represents the instantaneous location of the gravity wave ($U_{1}^{wave}=1 + 1/Fr$) and it coincides with the onset of the rise of the water level. The green dotted line (b) represents the front of the polymer advection ($U_{2}^{wave}=1 $) and it coincides with the location of the peak height of the water level. The black solid line (b) represents a traveling wave with the wave speed $U_{3}^{wave}=0.7$. The values of other parameters are $Fr=0.15$, $Pe \to \infty$, $B=0$, $DR_{max}=67\%$, and $l=2 \times 10^4$.}
\end{figure}

To gain insight into the basic characteristics of the dynamics of water height in the channel, first, we consider a tiny concentration of the polymer additives ($c_{in}=0.01$), which leads to linear dynamics. For the linear study, we ignore the polymeric dispersion and degradation. The polymer is being injected in the time interval $t=10\times 10^3-60\times 10^3$ (5.34-32 hours) and the injection concentration has been changed suddenly (Fig. \ref{cin_vs_t_Tinj20k_tnond.png}). The spatiotemporal dynamics of water level and polymeric concentration in the channel have been shown in Fig. \ref{hprime_vs_x_and_t} and Fig. \ref{cprime_vs_x_and_t}, respectively, and also (Supplementary Video-1).  During the polymeric injection, the height of the water level at any location in the channel initially increases, achieves a maximum value, and then decreases due to drag reduction. The dynamics of the water level can be characterized by three traveling waves, which are associated with (i) the onset of the rise of the water level, (ii) the peak height of the water level, and (iii) the onset of the decrease of water level below the base height, respectively. The injection of polymers in the channel creates a disturbance in the base flow and the front of the disturbance travels at the speed of the gravity wave. The speed of shallow water gravity wave in the downstream direction can be given as:
\begin{equation}\label{wave_speed}
U_{1}^{wave}=1 + 1/Fr.
\end{equation}
The onset of the rise of the water level in the channel coincides with the location of the gravity wave front (Fig. \ref{cprime_vs_x_and_t}). The speed of the gravity wave is much larger than the flow speed due to the small $Fr$. Hence, the water level starts to rise much before the polymer's arrival. The water level starts to decrease due to the drag reduction once the polymer arrives (Fig. \ref{cprime_vs_x_and_t}). Hence, the peak height of the water level coincides with the front of the polymer advection and the speed of the traveling wave associated with the peak water height is given by the bulk velocity of the fluid, which in the limit of small polymer concentration is barely perturbed: $U_{2}^{wave}=1$. The height of the water level decreases at a finite rate in the presence of polymers before it achieves a steady state. Therefore, before the water level goes below the base height and ultimately achieves a steady state, it remains higher than the base level for a significant amount of time even after the polymer's arrival (Fig. \ref{cprime_vs_x_and_t}). The maximum rise of water level increases downstream due to the cumulative effect of upstream disturbances. Therefore, the time duration within which the water level is higher than the base height increases downstream of the channel. The speed of the traveling wave associated with the onset of the decrease of water level below the base height is found to be $U_{3}^{wave} \approx 0.7$. This wave speed is independent of $Fr$, $c_{in}$, and $DR_{max}$. {The deviation of the onset of the decrease of water level below the base height from the wave speed $U_{3}^{wave}$ far downstream of the channel is due to the effect of the downstream boundary condition on subcritical flows.}
Once the injection of polymer is stopped (i.e., for $t>60 \times 10^3$ ), the water level temporarily further drops below the steady state height obtained during the polymeric injection before it bounces back to the base height.  This observation is complementary to the observed increase in height downstream of the polymer concentration front after the start of injection. 

\subsection{Downstream overshoot mechanism}

\begin{figure}[!ht]
\centering
\begin{subfigure}[b]{0.48\textwidth} 
\includegraphics[width=\textwidth]{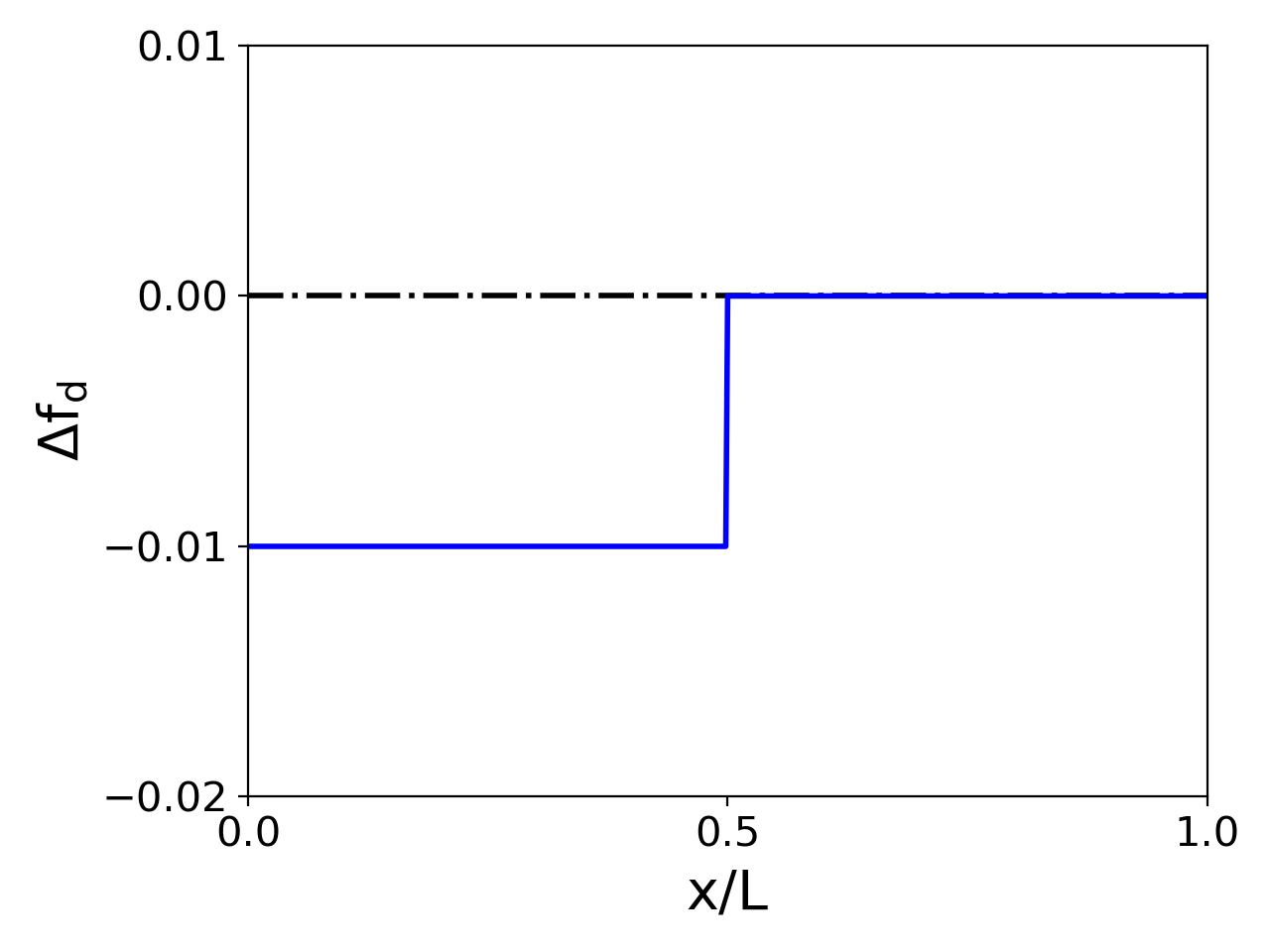}
\caption{}
\label{delta_fd_vs_x}
\end{subfigure}
\begin{subfigure}[b]{.48\textwidth}
\includegraphics[width=\textwidth]{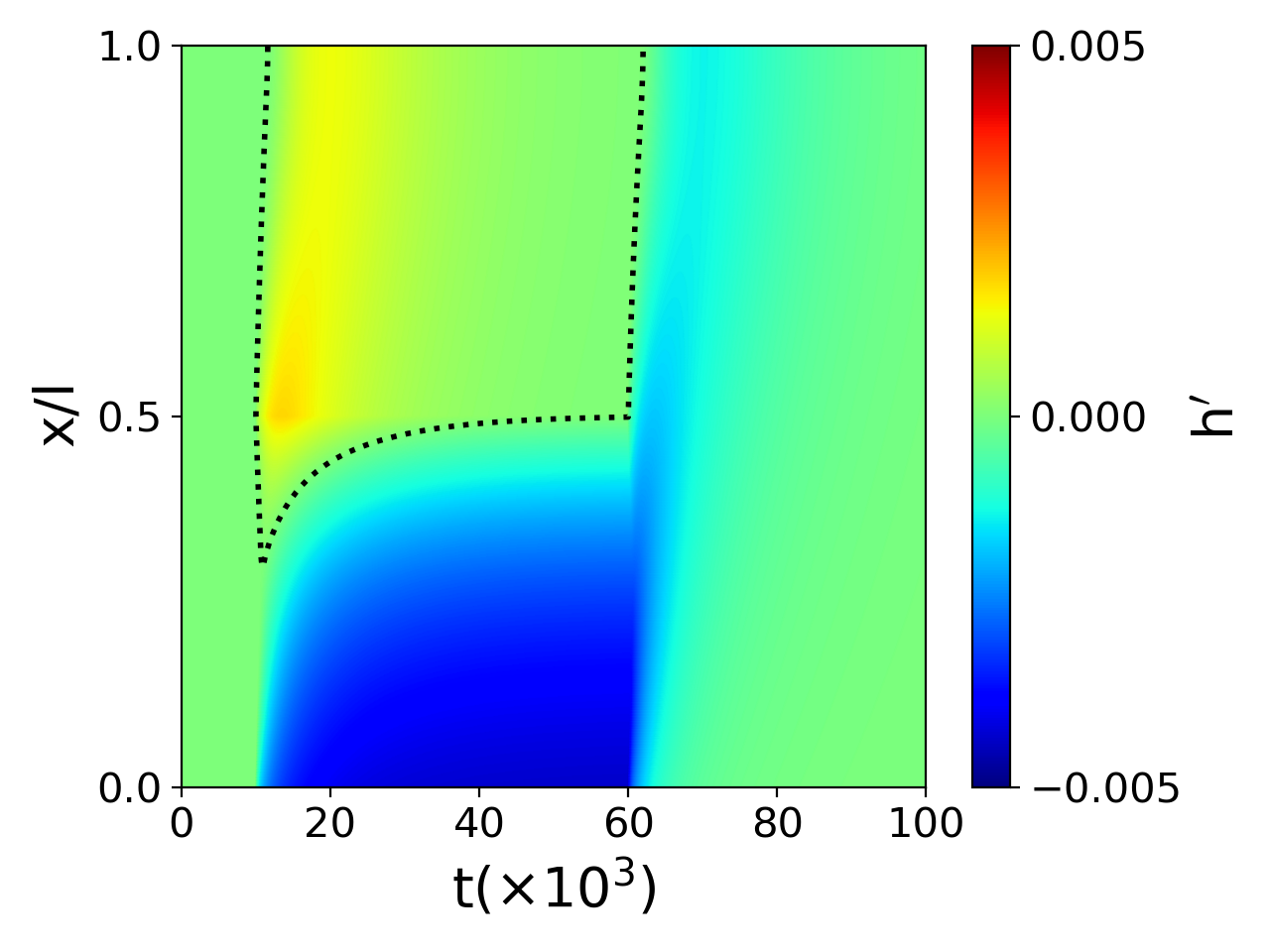}
\caption{}
\label{hprime_vs_x_and_t_step_fd}
\end{subfigure}
\caption{(a) A sudden change in friction in the front half of the channel in the time interval $t=10 \times 10^3-60\times 10^3$. (b) Spatiotemporal contour of height perturbation for a sudden change in friction upstream of the channel as shown in (a). Other parameters are $Fr=0.15$ and $l=2 \times 10^4$.}
\label{step_fd_change}
\end{figure}

The increase in water height downstream of polymer injection arises from fluid inertia and the sudden drop in friction due to polymer injection.  To illustrate this mechanism, we performed simulations where we reduce friction in the front half of the channel as shown in Fig. \ref{delta_fd_vs_x} in the time interval $t=10 \times 10^3-60\times 10^3$ and analyze water height in the channel (Fig. \ref{hprime_vs_x_and_t_step_fd}). The sudden reduction in friction in the upstream half of the channel leads to enhanced flow speed and hence fluid momentum, which induces a height overshoot in the downstream half of the channel. The height overshoots last for a long time, as its natural decay is slow (Fig. \ref{hprime_vs_x_and_t_step_fd}). However, the overshoot in the channel having polymeric injection (Fig. \ref{hprime_vs_x_and_t}) decays relatively faster because the polymeric advection reduces the drag at the overshoot location which amplifies the overshoot decay rate. Thus, the interplay between fluid inertia and drag reduction due to polymer additives leads to the water height overshoot in the channel. Once the friction in the upstream half of the channel is brought back to the original value at $t=60\times 10^3$, fluid momentum decreases in the upstream region, which creates a temporary water height undershoot in the downstream half of the channel before the water height bounces back to the base level. This is why we see a water height undershoot downstream of the channel once the polymeric injection is stopped (Fig. \ref{hprime_vs_x_and_t}).

\subsection{Nonlinear dynamics}

\begin{figure}[!ht]
\centering
\begin{subfigure}[b]{0.48\textwidth} 
\includegraphics[width=\textwidth]{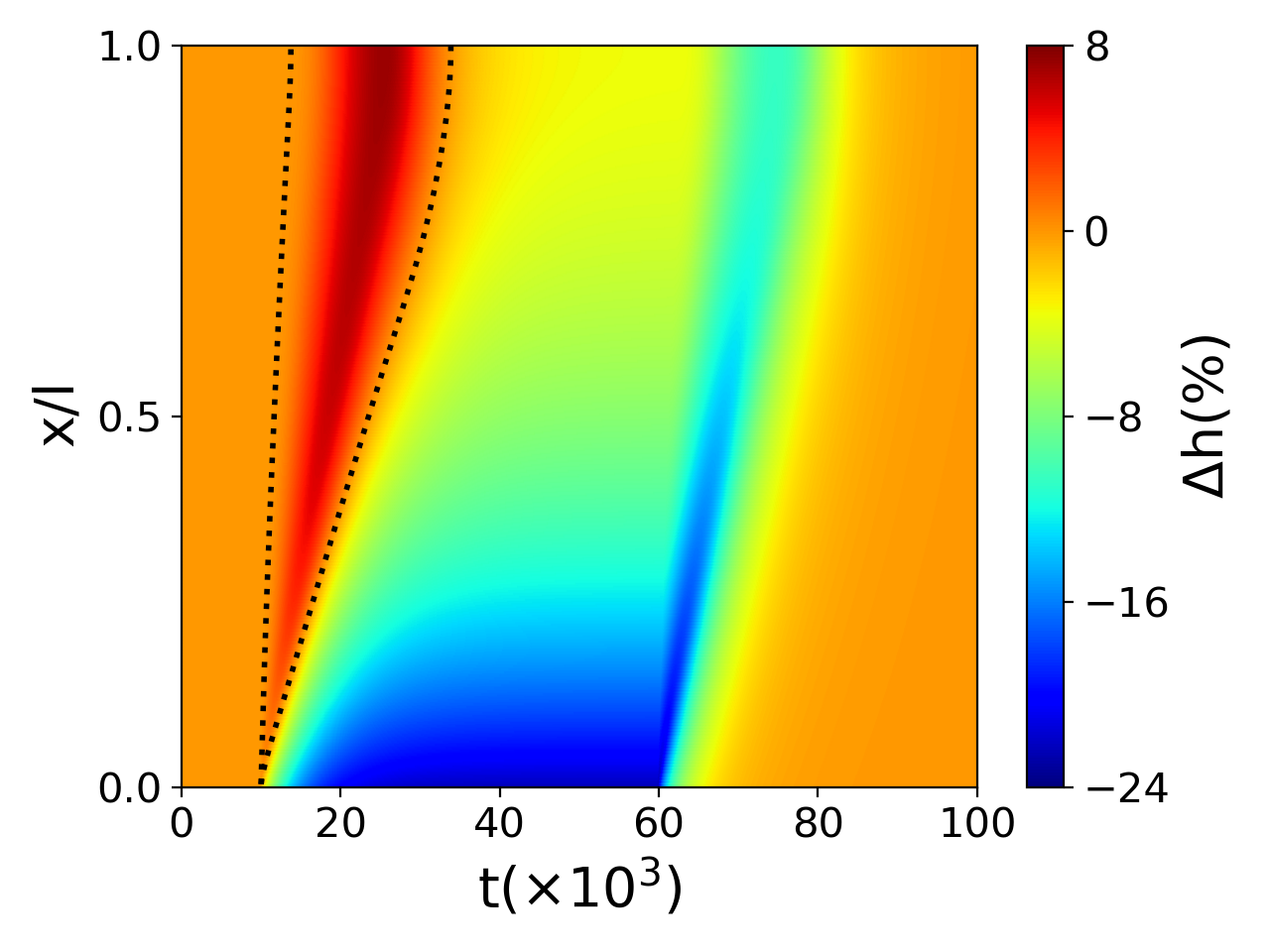}
\caption{}
\label{density_h_h20k_Fr015_S1e_4_Pe1e_2_B1e_4_mdr3_Exp1e_2_dotted.png}
\end{subfigure}
\begin{subfigure}[b]{.48\textwidth}
\includegraphics[width=\textwidth]{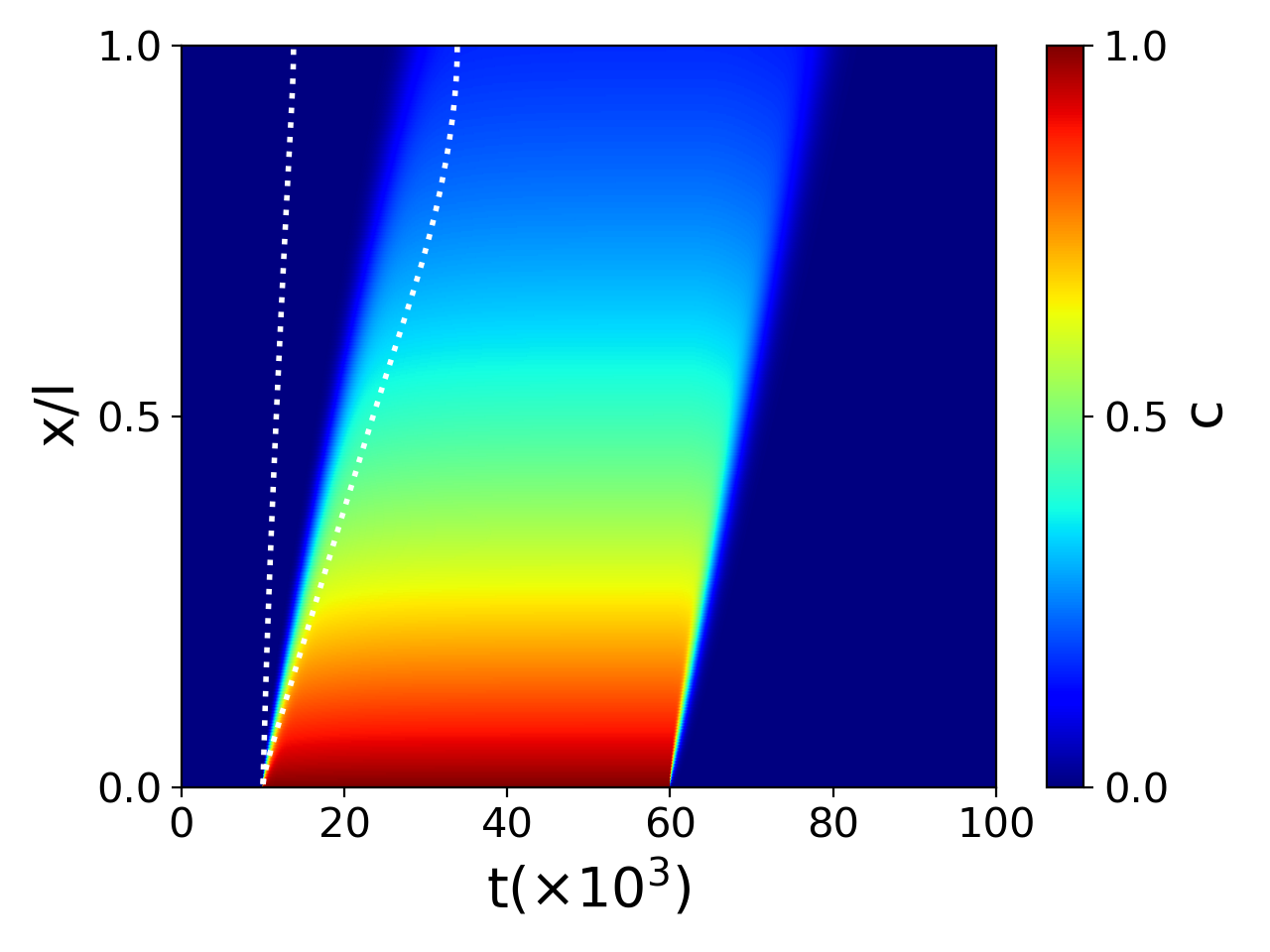}
\caption{}
\label{density_c_h20k_Fr015_S1e_4_Pe1e_2_B1e_4_mdr3_Exp1e_2_hpos.png}
\end{subfigure}
\begin{subfigure}[b]{0.48\textwidth} 
\includegraphics[width=\textwidth]{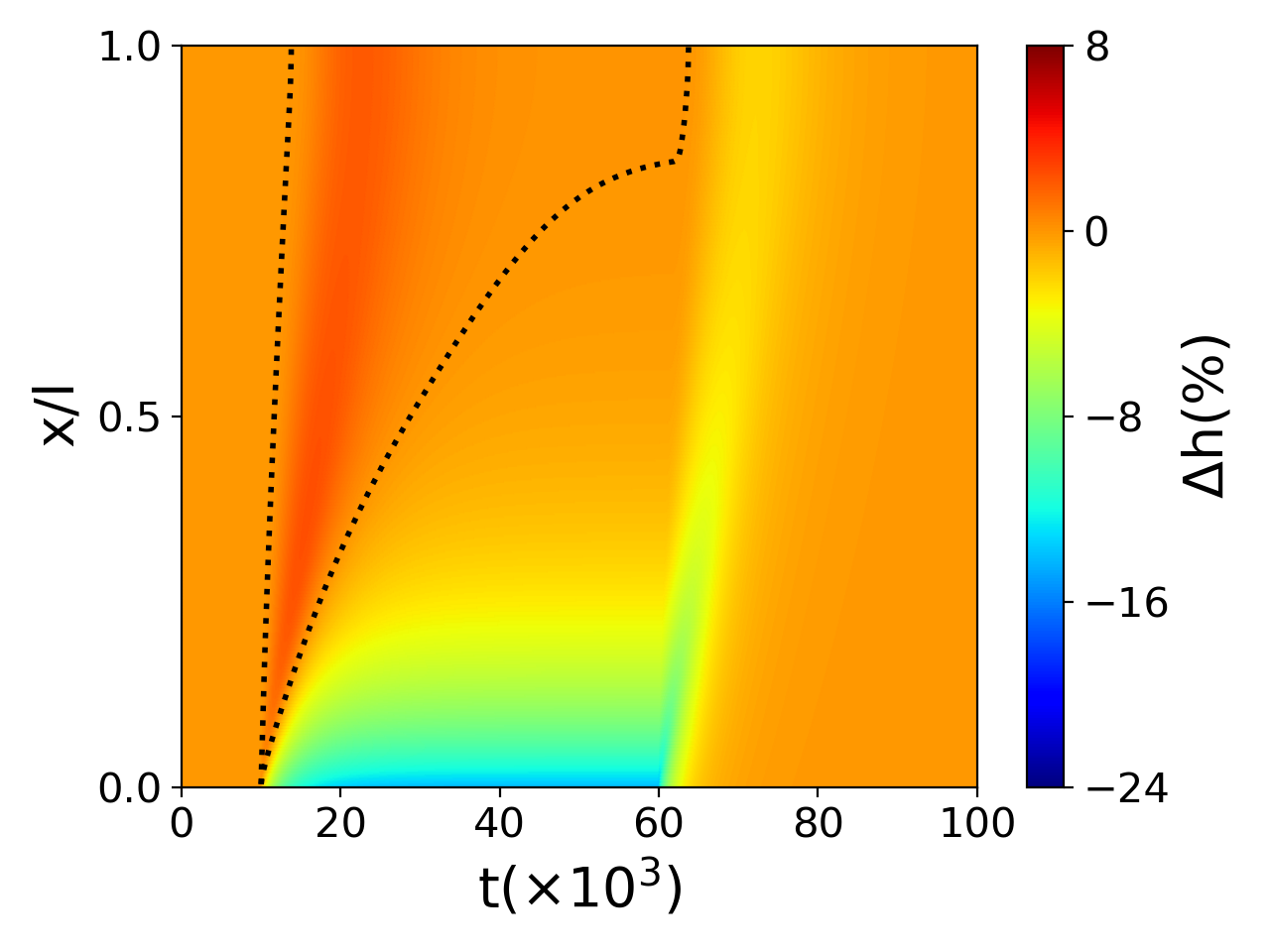}
\caption{}
\label{density_h_h20k_Fr015_S1e_4_Pe1e_2_B3e_4_mdr3_Exp1e_2_dotted.png}
\end{subfigure}
\begin{subfigure}[b]{.48\textwidth}
\includegraphics[width=\textwidth]{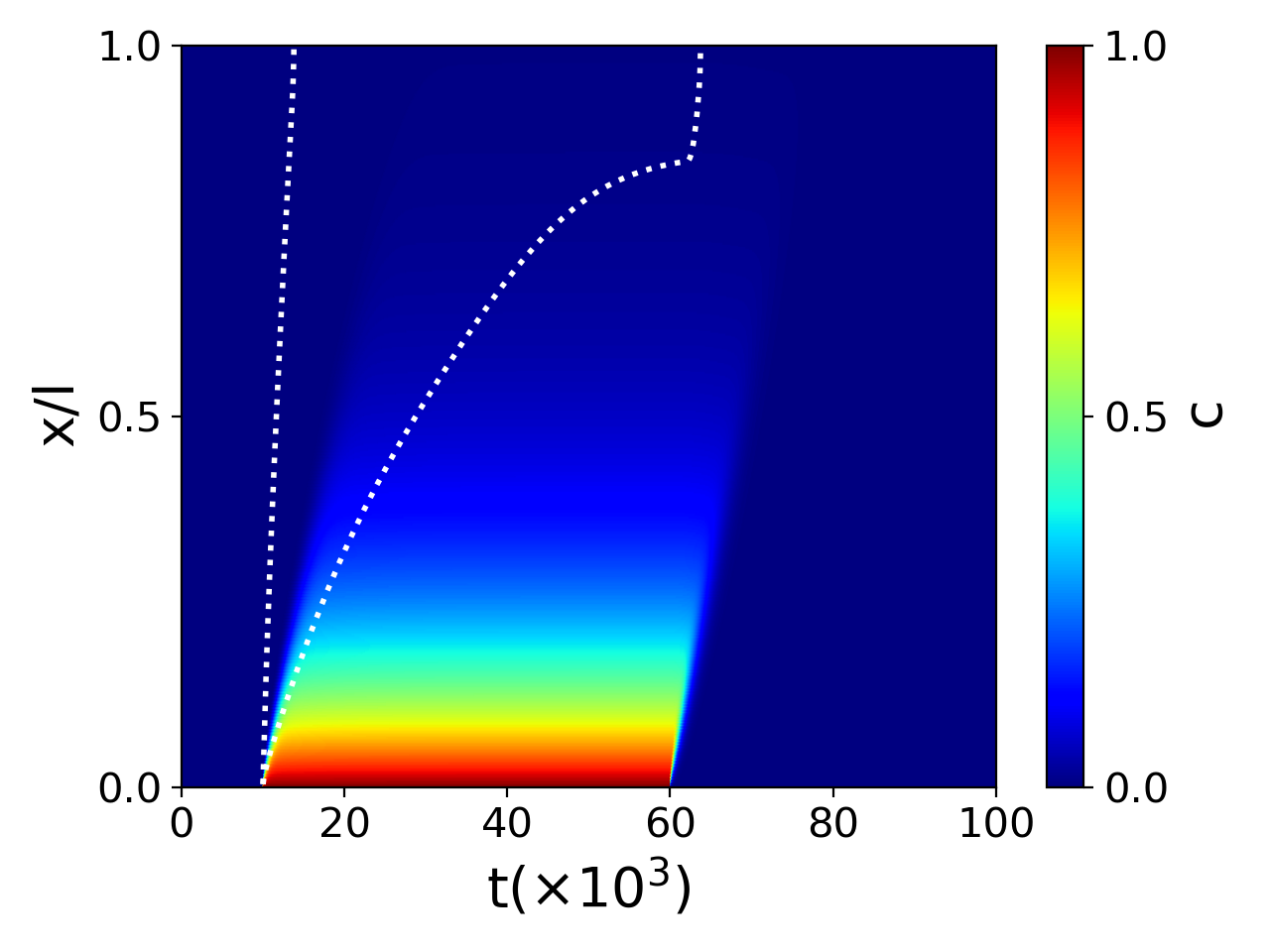}
\caption{}
\label{density_c_h20k_Fr015_S1e_4_Pe1e_2_B3e_4_mdr3_Exp1e_2_hpos.png}
\end{subfigure}
\caption{Spatiotemporal contours of (a,c) the change in water height ($\Delta h$) and (b,d) polymeric concentration for the degradation rate (a,b) $B=1 \times 10^{-4}$ and (c,d) $B=3 \times 10^{-4}$. The space enclosed by dotted lines represents the region where the water level is higher than the base height during the polymer injection. The values of other parameters are $Fr=0.15$, $Pe = 10^{-2}$, $DR_{max}=67\%$, and $l=2 \times 10^4$.}
\label{height_concentration_with_degradation}
\end{figure}

The investigation of linearized equations in the absence of polymeric degradation and dispersion reveals three distinct traveling waves due to gravity wave, polymer advection, and drag reduction (Fig. \ref{cprime_vs_x_and_t}). Next, we investigate the nonlinear dynamics of water height in the channel in the presence of polymeric degradation and dispersion (Supplementary Video-2). Similar to the linear dynamics, the water level at any location in the channel initially rises, achieves a maximum value, and then decreases (Fig. \ref{height_concentration_with_degradation}). The rate of decline of water level and the steady state height of water depend on local polymeric concentration. Therefore, the degradation of polymeric chains plays a very important role during the retraction of water level after achieving the maximum height (Figs. \ref{density_h_h20k_Fr015_S1e_4_Pe1e_2_B3e_4_mdr3_Exp1e_2_dotted.png} and \ref{density_c_h20k_Fr015_S1e_4_Pe1e_2_B3e_4_mdr3_Exp1e_2_hpos.png}). Due to degradation, the polymeric concentration decreases downstream leading to a slower decay of the water level. Therefore, the time duration in which the water level at any location in the channel remains higher than the base height increases as the polymeric degradation rate increases. Further, the duration of height overshoot increases downstream of the channel more rapidly as the degradation rate increases (Figs. \ref{density_h_h20k_Fr015_S1e_4_Pe1e_2_B1e_4_mdr3_Exp1e_2_dotted.png} and \ref{density_h_h20k_Fr015_S1e_4_Pe1e_2_B3e_4_mdr3_Exp1e_2_dotted.png}). For a sufficiently large degradation rate, the polymeric concentration far downstream becomes negligible (Fig. \ref{density_c_h20k_Fr015_S1e_4_Pe1e_2_B3e_4_mdr3_Exp1e_2_hpos.png}). In such a situation, the water level far downstream of the channel remains higher than the base height throughout the experiments (Fig. \ref{density_h_h20k_Fr015_S1e_4_Pe1e_2_B3e_4_mdr3_Exp1e_2_dotted.png}).

\begin{figure}[!ht]
\centering
\begin{subfigure}[b]{0.48\textwidth} 
\includegraphics[width=\textwidth]{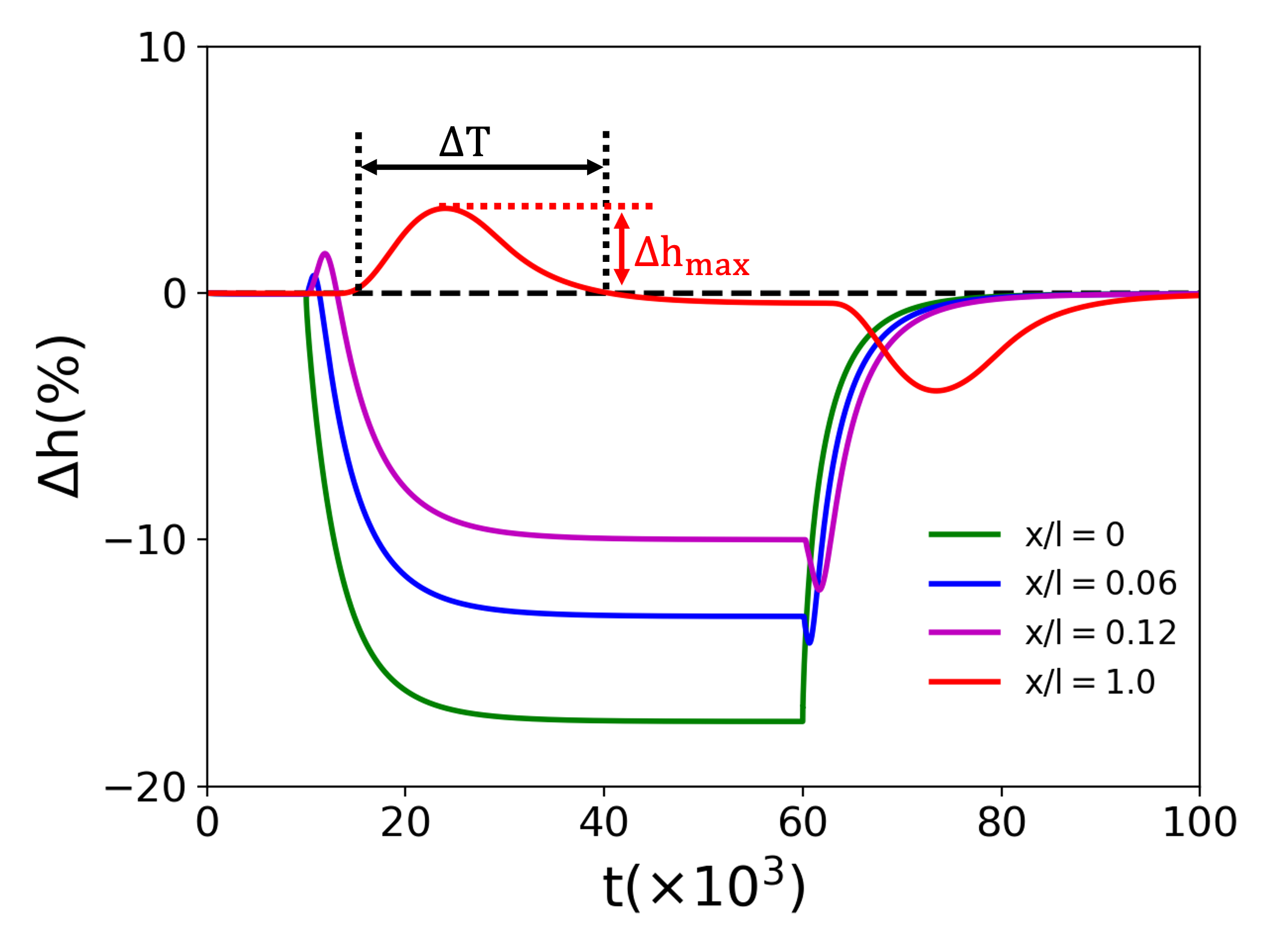}
\caption{}
\label{height_diff_vs_t_h20k_Fr015_S1e_4_Pe1e_2_B1e_4_mdr3_Exp1e_2_modified.png}
\end{subfigure}
\begin{subfigure}[b]{.48\textwidth}
\includegraphics[width=\textwidth]{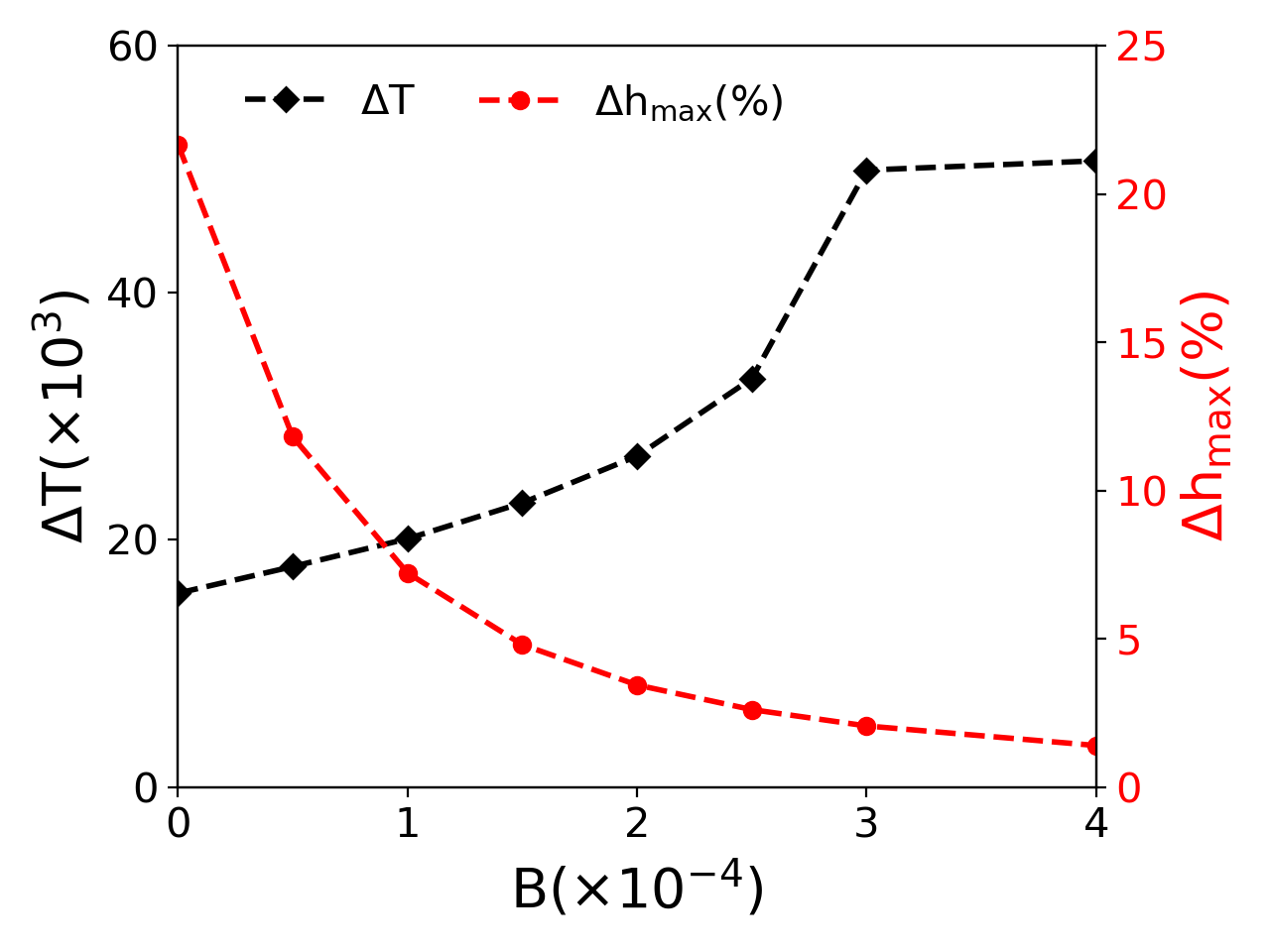}
\caption{}
\label{dt_dhmax_vs_B_h20k_Fr015_S1e_4_Pe1e_2_mdr3_Exp1e_2.png}
\end{subfigure}
\begin{subfigure}[b]{0.48\textwidth} 
\includegraphics[width=\textwidth]{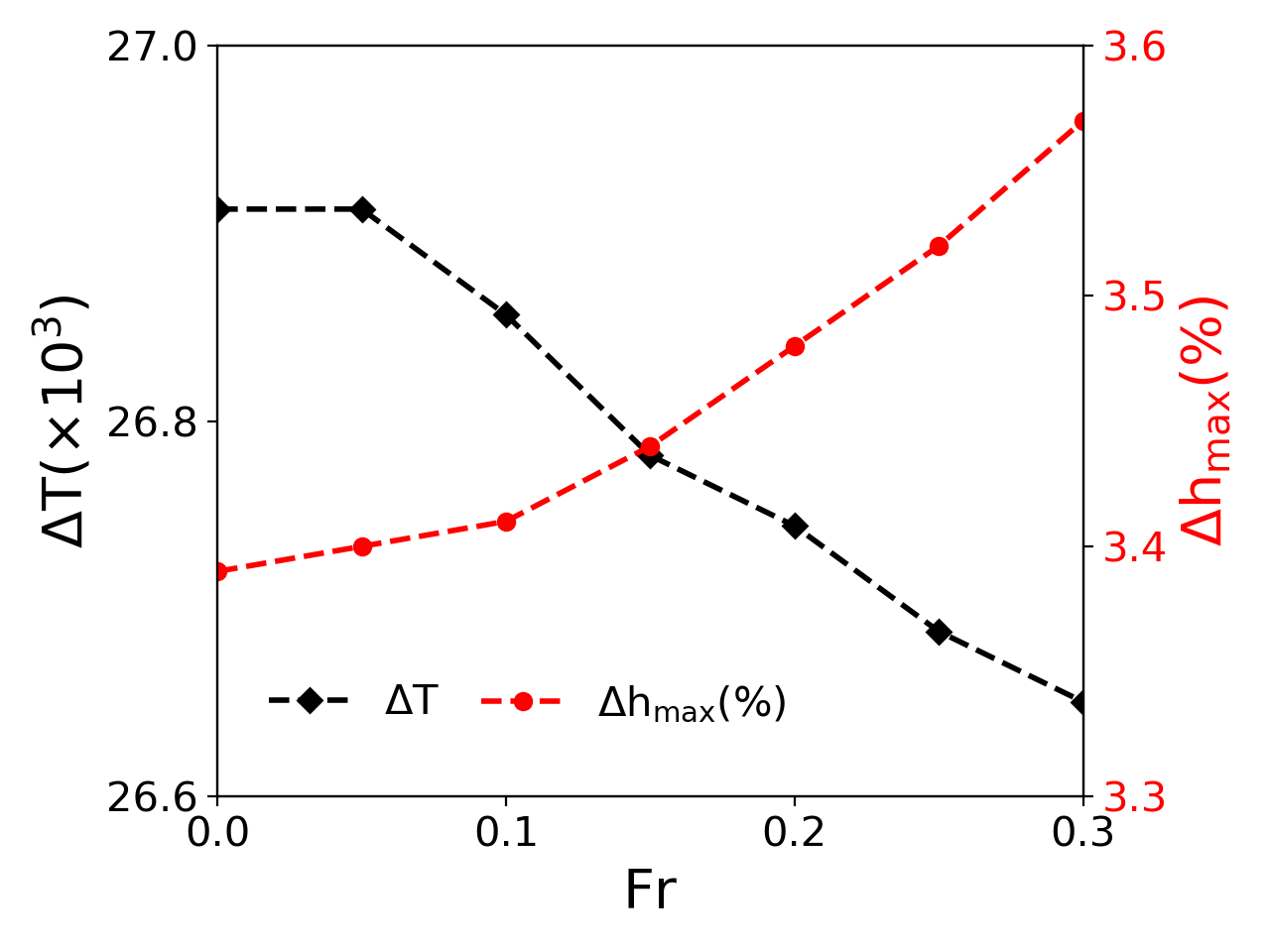}
\caption{}
\label{dt_dhmax_vs_Fr_L20k_S1e_4_Pe1e_2_mdr3_Exp1e_2_B2e_4.png}
\end{subfigure}
\begin{subfigure}[b]{.48\textwidth}
\includegraphics[width=\textwidth]{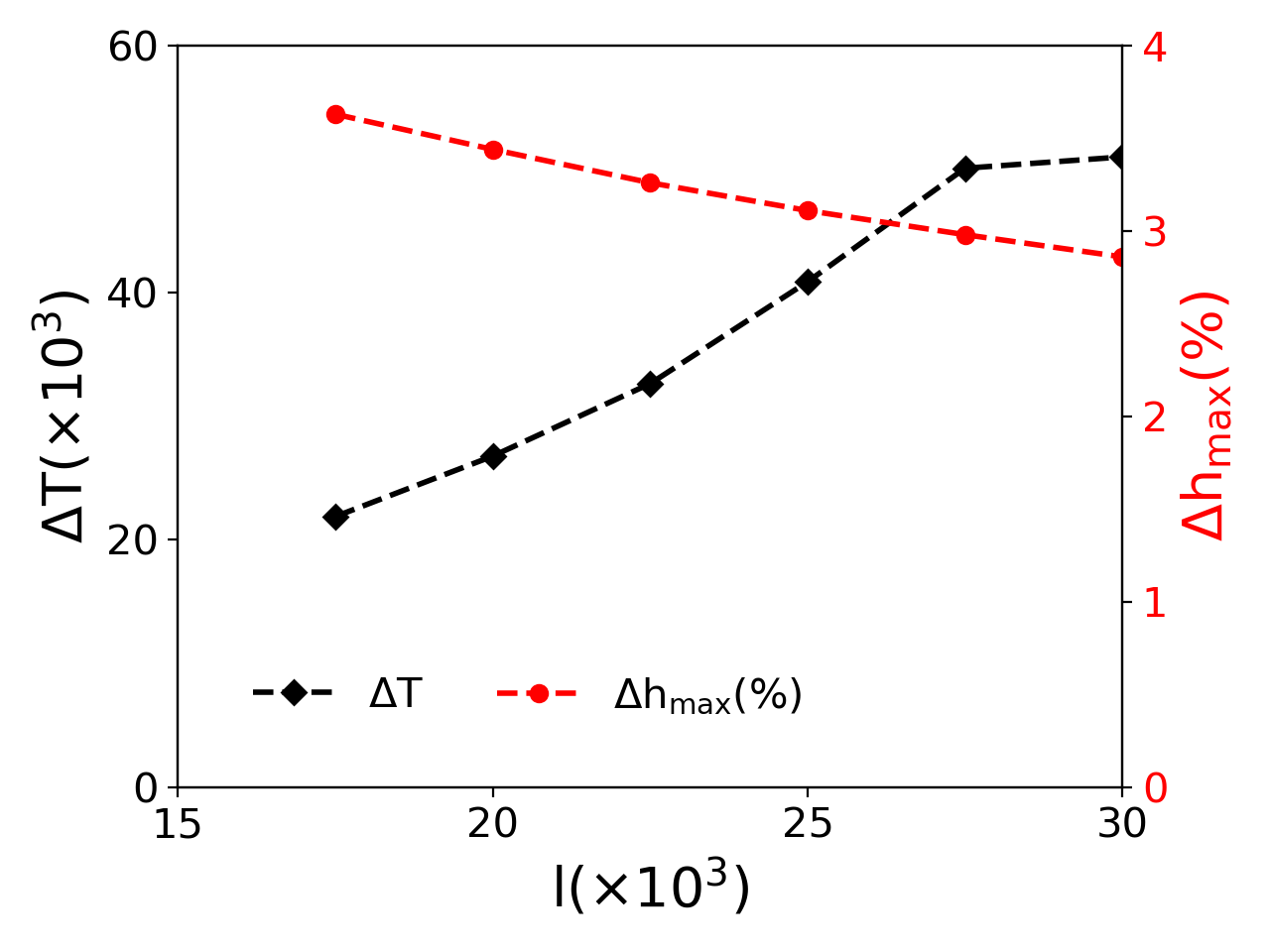}
\caption{}
\label{dt_dhmax_vs_L_Fr015_S1e_4_Pe1e_2_mdr3_Exp1e_2_B2e_4.png}
\end{subfigure}
\caption{ (a) Change in water level as a function of time at a few specific locations along the length of the channel for $B=2 \times 10^{-4}$, $Fr=0.15$, and $l=2 \times 10^4$. The time duration in which the water level is higher than the base height ($\Delta T$) and the maximum rise of water level ($\Delta h_{max}$) at the exit of the channel for different values of (b) degradation rate ($B$) at $Fr=0.15$ and $l=2 \times 10^4$, (c) Froude number ($Fr$) at $B=2 \times 10^{-4}$ and $l=2 \times 10^4$, and (d) channel length ($l$) at $Fr=0.15$ and $B=2 \times 10^{-4}$. Other parameters are $Pe = 10^{-2}$ and $DR_{max}=67\%$.}
\end{figure}

To further explore the temporal evolution of water height at a given location in the channel, we plot the time-dependent water level at a few specific locations along the length of the channel at an intermediate degradation rate (Fig. \ref{height_diff_vs_t_h20k_Fr015_S1e_4_Pe1e_2_B1e_4_mdr3_Exp1e_2_modified.png}). Close to the injection point, the water level quickly achieves a steady state at a height lower than the base level soon after an initial height overshoot. However, the overshoot downstream of the channel lasts for a long time ($\sim 26 \times 10^3 $(14 hrs)) and the steady state height is marginally below the base height (Fig. \ref{height_diff_vs_t_h20k_Fr015_S1e_4_Pe1e_2_B1e_4_mdr3_Exp1e_2_modified.png}). A similar water level overshoot has been reported in the field experiment performed in the irrigation canal \cite{Bouchenafa2021}. Further, we quantify the time-duration of overshoot ($\Delta T$) and also the maximum height achieved during this overshoot ($\Delta h_{max}$) at the exit of the channel (Fig. \ref{height_diff_vs_t_h20k_Fr015_S1e_4_Pe1e_2_B1e_4_mdr3_Exp1e_2_modified.png}). The overshoot duration ($\Delta T$) increases monotonically with polymeric degradation rate ($B$) until it saturates for higher degradation rates (Fig. \ref{dt_dhmax_vs_B_h20k_Fr015_S1e_4_Pe1e_2_mdr3_Exp1e_2.png}). The concentration of polymeric additives at the exit decreases as the degradation rate increases, leading to a slower decline of water level and hence an elevated duration of overshoot. At a sufficiently large degradation rate ($B>2.5 \times 10^{-4}$), the rate of height decline is small enough due to insignificant polymeric concentration that the height overshoot lasts longer than the duration of polymer injection. Therefore, the water level remains higher than the base height throughout the experiment, and the value of $\Delta T$ saturates. It is worth noting that the overshoot lasts for $\sim 15 \times 10^3$ time unit ($\sim8$ hours) even in the absence of the polymeric degradation and it gets longer with polymeric degradation. The maximum height obtained during the overshoot ($\Delta h_{max}$) decreases exponentially with the polymeric degradation rate. The degradation of polymer leads to an alleviated drag reduction, which creates a less severe height overshoot downstream of the channel. Despite the decrease of $\Delta h_{max}$ with $B$, the overshoot time $\Delta T$ increases with $B$ because the drop in the height decline rate is more pronounced compared to the drop in $\Delta h_{max}$. The effect of the Froude number on the overshoot has been shown in Fig. \ref{dt_dhmax_vs_Fr_L20k_S1e_4_Pe1e_2_mdr3_Exp1e_2_B2e_4.png}. The gravity wave travels slower as $Fr$ increases (Eq. \ref{wave_speed}), which delays the onset of the rise of the water level. Therefore, the overshoot time decreases with increasing $Fr$. The maximum overshoot height increases with $Fr$ due to the enhanced effect of inertia at large $Fr$. However, the overall influence of $Fr$ on $\Delta T$ and $\Delta h_{max}$ for the range of $Fr$ relevant in the present study is marginal because $Fr \ll 1$. The effect of channel length on the overshoot has trends similar to the polymeric degradation rate (Fig. \ref{dt_dhmax_vs_L_Fr015_S1e_4_Pe1e_2_mdr3_Exp1e_2_B2e_4.png}). For a given degradation rate, the remaining polymer (i.e., polymeric concentration) toward the exit of the channel decreases with the increasing length of the channel. Therefore, the overshoot time ($\Delta T$) increases and the peak height of the overshoot ($\Delta h_{max}$) decreases for an increasing value of channel length.

\begin{figure}[!ht]
\centering
\begin{subfigure}[b]{0.48\textwidth} 
\includegraphics[width=\textwidth]{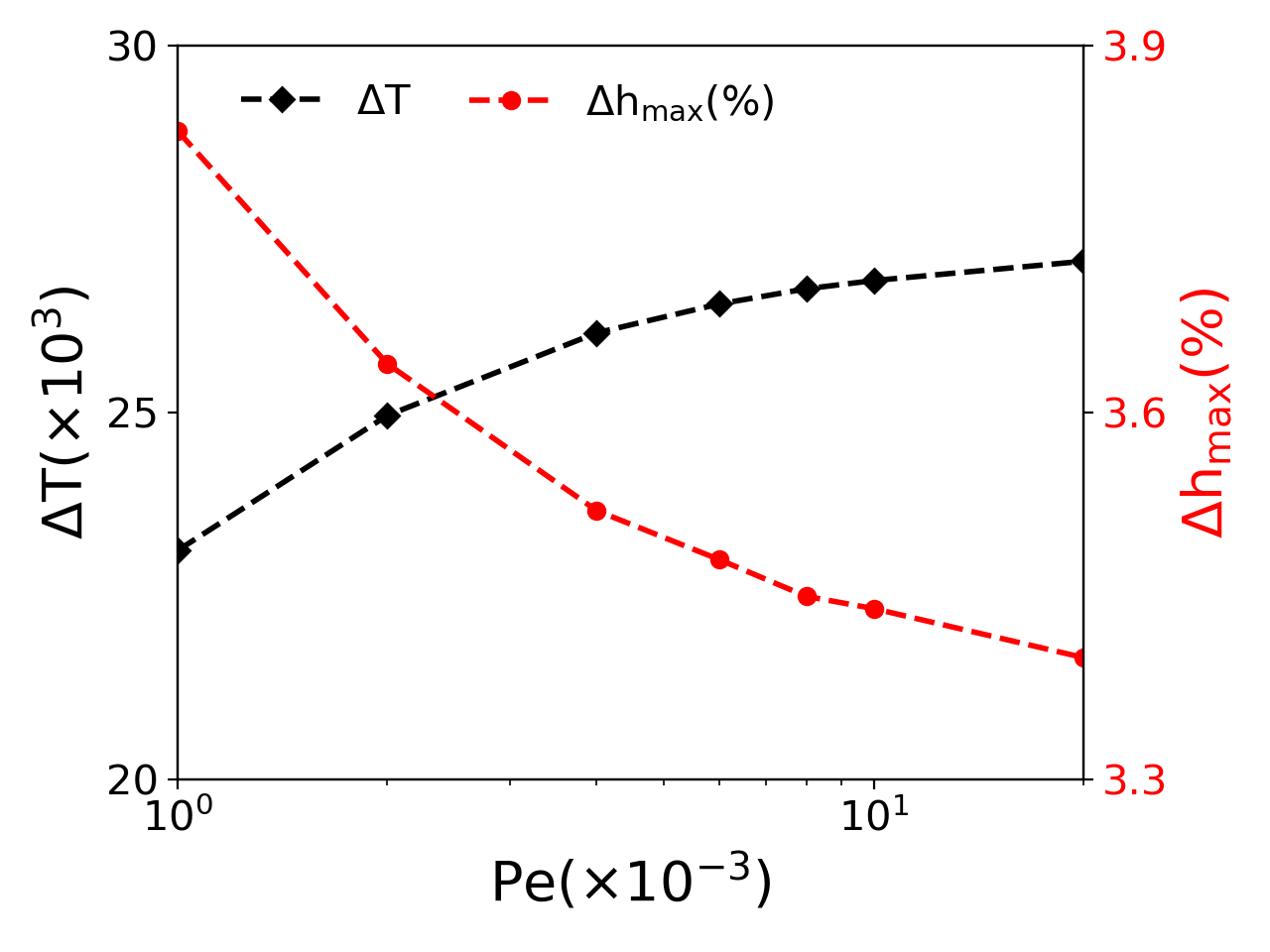}
\caption{}
\label{dt_dhmax_vs_Pe_L20k_Fr015_S1e_4_mdr3_Exp1e_2_B2e_4_semilogx.png}
\end{subfigure}
\begin{subfigure}[b]{.48\textwidth}
\includegraphics[width=\textwidth]{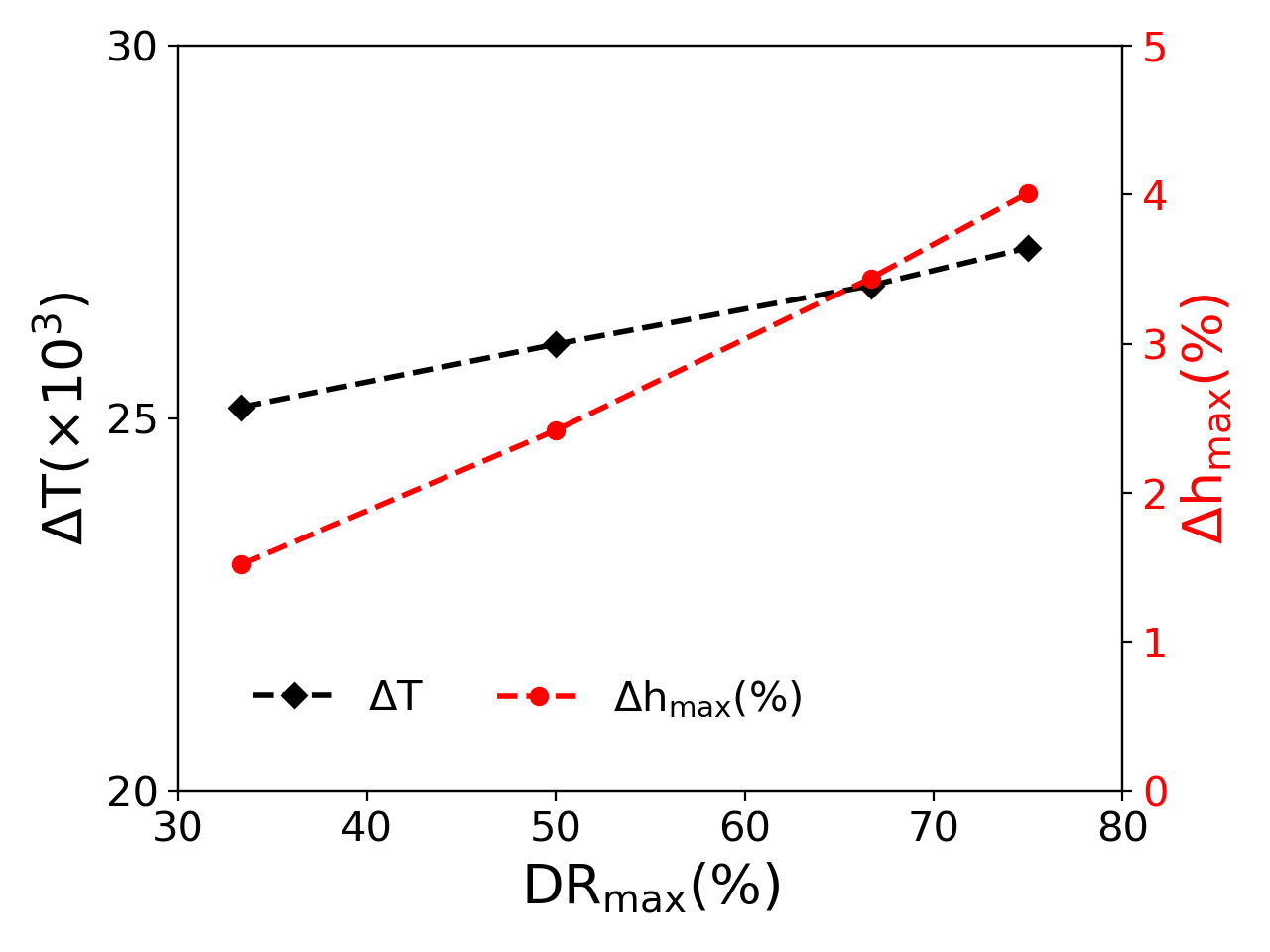}
\caption{}
\label{dt_dhmax_vs_MDR_h20k_Fr015_S1e_4_Pe1e_2_B2e_4.png}
\end{subfigure}
\caption{ Time duration of overshoot ($\Delta T$) and the maximum rise of water level during overshoot ($\Delta h_{max}$) at the exit of the channel for different values of (a) Péclet number ($Pe$) at $DR_{max}=67\%$ and (b) maximum drag reduction ($DR_{max}$) at $Pe = 10^{-2}$. The values of other parameters are $Fr=0.15$, $B=2 \times 10^{-4}$, and $l=2 \times 10^4$.}
\end{figure}

The effect of turbulent dispersion on the overshoot of the water level has been shown in Fig. \ref{dt_dhmax_vs_Pe_L20k_Fr015_S1e_4_mdr3_Exp1e_2_B2e_4_semilogx.png}. The dispersion becomes stronger with a decreasing value of $Pe$. Due to dispersion, the polymer arrives in advance compared to pure advection at any location downstream. Therefore, the water level starts to decline earlier as turbulent dispersion increases, which leads to the decrease of overshoot time ($\Delta T$) as $Pe$ decreases. Polymeric dispersion also mitigates the effect of degradation resulting in a relatively higher polymeric concentration at any location downstream as dispersion increases. This is the reason behind the increase of the maximum water height during overshoot as $Pe$ decreases. 

The maximum drag reduction ($DR_{max}$) due to polymer additives varies significantly depending on the chemistry of polymeric molecules and the physical setup of the experiment \cite{Virk1975,Virk1997,Han2017,Bhambri2016}. The characteristics of water overshoot ($\Delta T$ and $\Delta h_{max}$) for different values of $DR_{max}$ have been shown in Fig. \ref{dt_dhmax_vs_MDR_h20k_Fr015_S1e_4_Pe1e_2_B2e_4.png}. For a larger $DR_{max}$, the change in fluid momentum due to the injection of polymer is larger. Therefore, the height overshoot at the exit becomes more severe as $DR_{max}$ increases, which leads to the increase of $\Delta h_{max}$ with $DR_{max}$. This also leads to the increase of overshoot duration ($\Delta T$) with $DR_{max}$. However, $\Delta T$ increases slowly with $DR_{max}$ compared to $\Delta h_{max}$, because the rate of decline of water level also increases with $DR_{max}$ during the overshoot.

\begin{figure}[!ht]
 \centering
 \includegraphics[width=.5\textwidth]{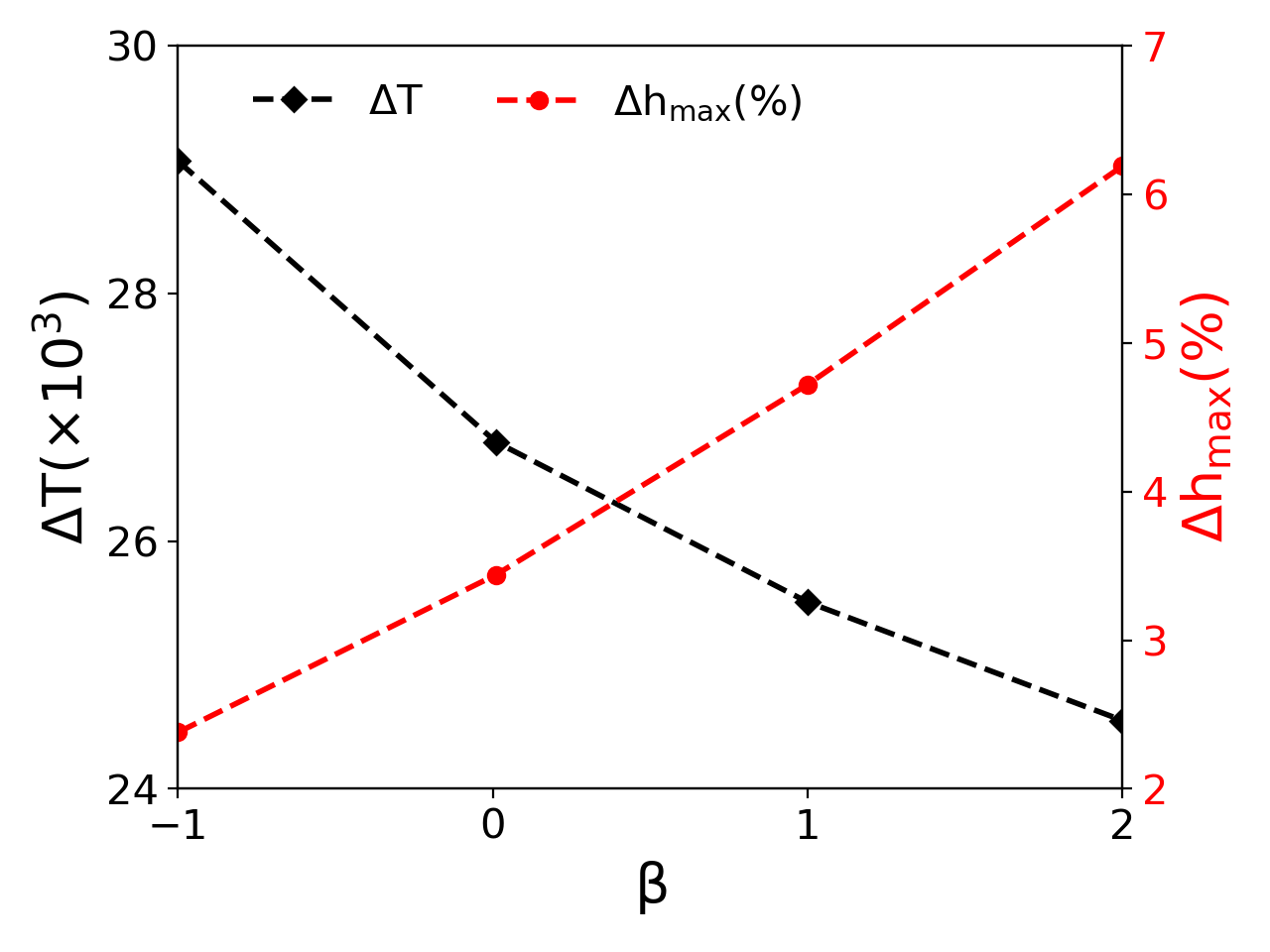}
 \caption{\label{dt_dhmax_vs_beta_L20k_Fr015_S1e_4_Pe1e_2_mdr3_Exp_B2e_4.png} Time duration of overshoot ($\Delta T$) and the maximum rise of water level during overshoot ($\Delta h_{max}$) at the exit of the channel for different profiles of $f_d$. The values of other parameters are $Fr=0.15$, $Pe = 10^{-2}$, $B=2 \times 10^{-4}$, $DR_{max}=67\%$, and $l=2 \times 10^4$.}
 \end{figure}

We also investigate the effect of the nonlinear profiles of concentration-dependent drag ($f_d$) on water level dynamics as $f_d$ profile of polymeric solution is often nonlinear (Fig. \ref{f_d_vs_c.png}). The maximum water height during the overshoot ($\Delta h_{max}$) increases monotonically as the convexity of the $f_d$ profile decreases (Fig. \ref{dt_dhmax_vs_beta_L20k_Fr015_S1e_4_Pe1e_2_mdr3_Exp_B2e_4.png}). The local drag inside the channel decreases with the decreasing convexity of $f_d$ profile, which induces an enhanced disturbance due to polymer injection and leads to an elevated $\Delta h_{max}$ with the increasing value of $\beta$. The rate of the decline of water level increases as the convexity of $f_d$ profile decreases due to lower local drag. The enhanced decline rate of water level overcomes the increase of $\Delta h_{max}$ with $\beta$. Therefore, the overshoot duration $\Delta T$ decreases monotonically as the convexity of $f_d$ profile decreases (Fig. \ref{dt_dhmax_vs_beta_L20k_Fr015_S1e_4_Pe1e_2_mdr3_Exp_B2e_4.png}).  

\begin{figure}[!ht]
\centering
\begin{subfigure}[b]{.32\textwidth}
\includegraphics[width=\textwidth]{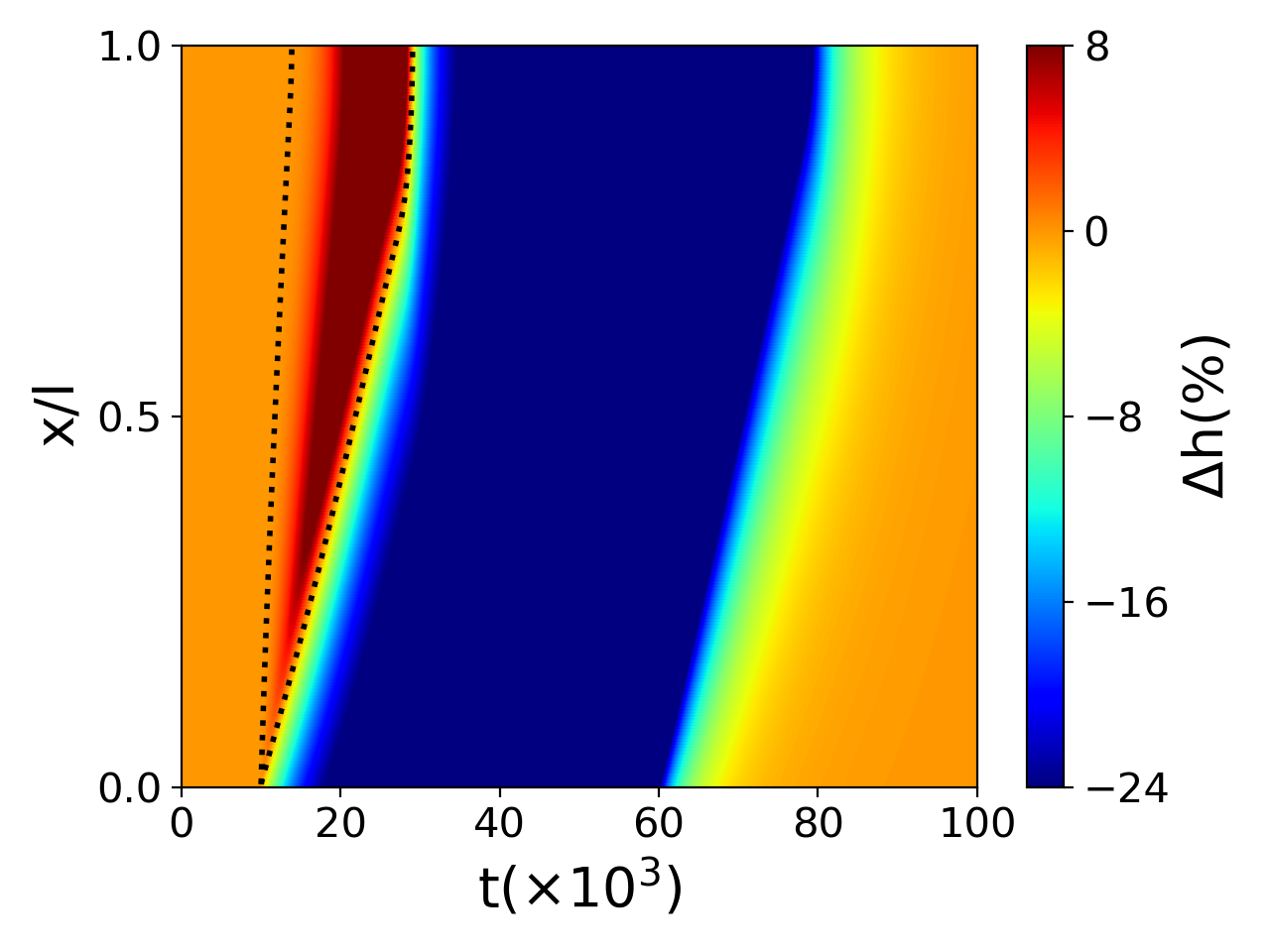}
\caption{}
\label{density_h_h20k_Fr015_S1e_4_Pe1e10_B0e_4_mdr3_Exp1e_2.png}
\end{subfigure}
\begin{subfigure}[b]{.32\textwidth}
\includegraphics[width=\textwidth]{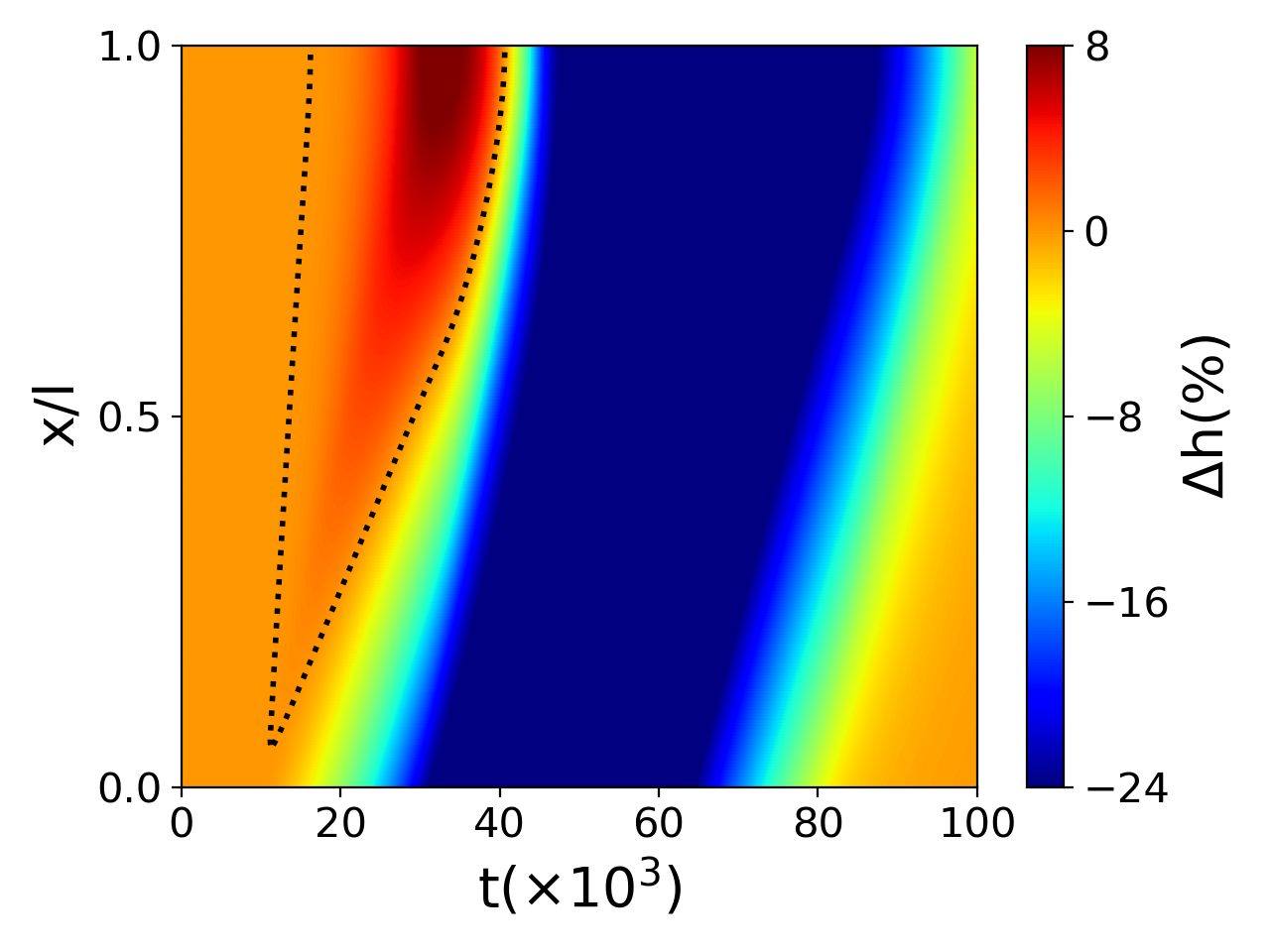}
\caption{}
\label{density_h_h20k_Fr015_S1e_4_Pe1e10_B0e_4_mdr3_Exp1e_2_Tinj20k.png}
\end{subfigure}
\begin{subfigure}[b]{.32\textwidth}
\includegraphics[width=\textwidth]{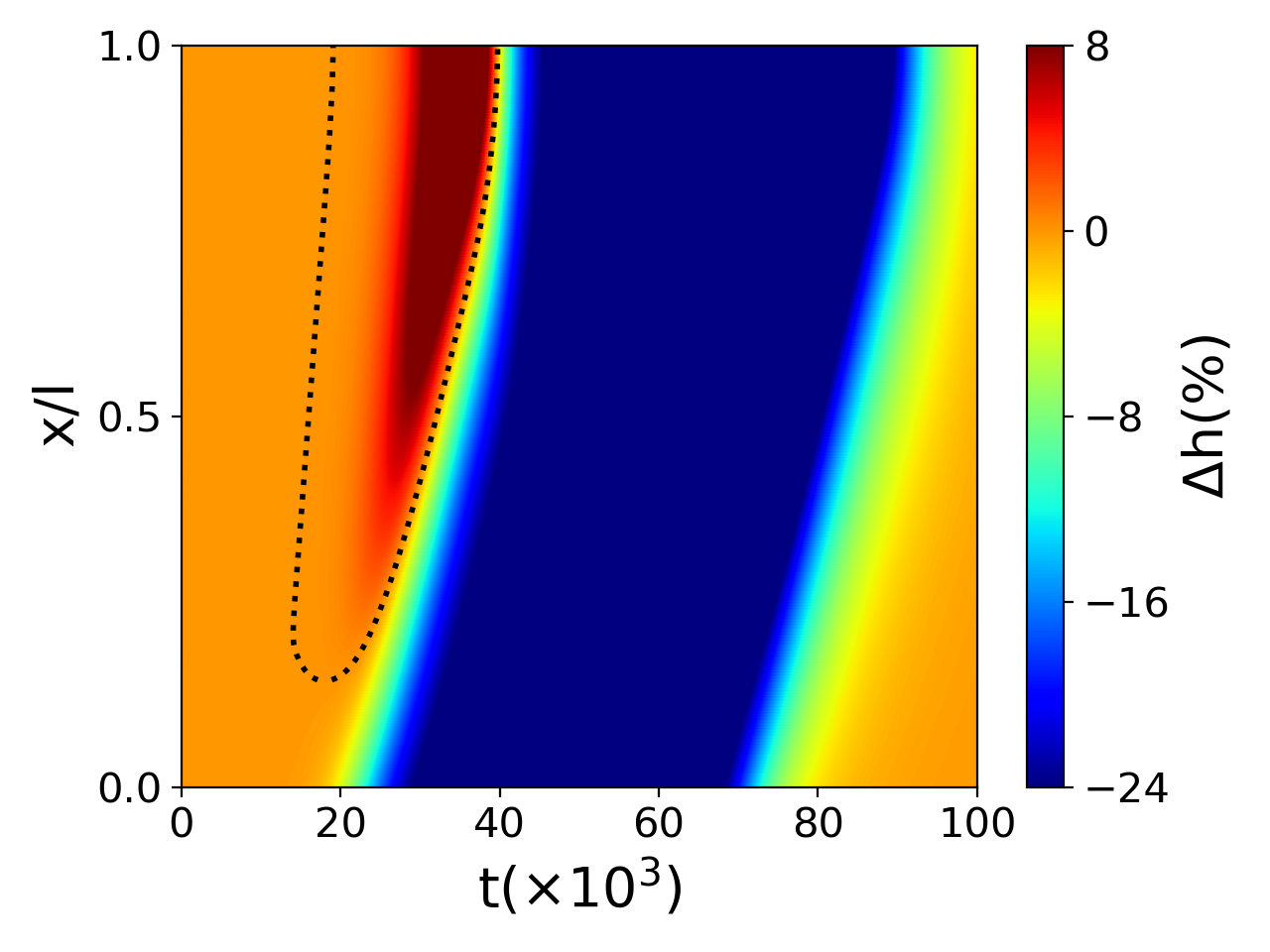}
\caption{}
\label{density_h_h20k_Fr015_S1e_4_Pe1e10_B0e_4_mdr3_Exp1e_2_Tinj20k_tanh.png}
\end{subfigure}
\begin{subfigure}[b]{.32\textwidth}
\includegraphics[width=\textwidth]{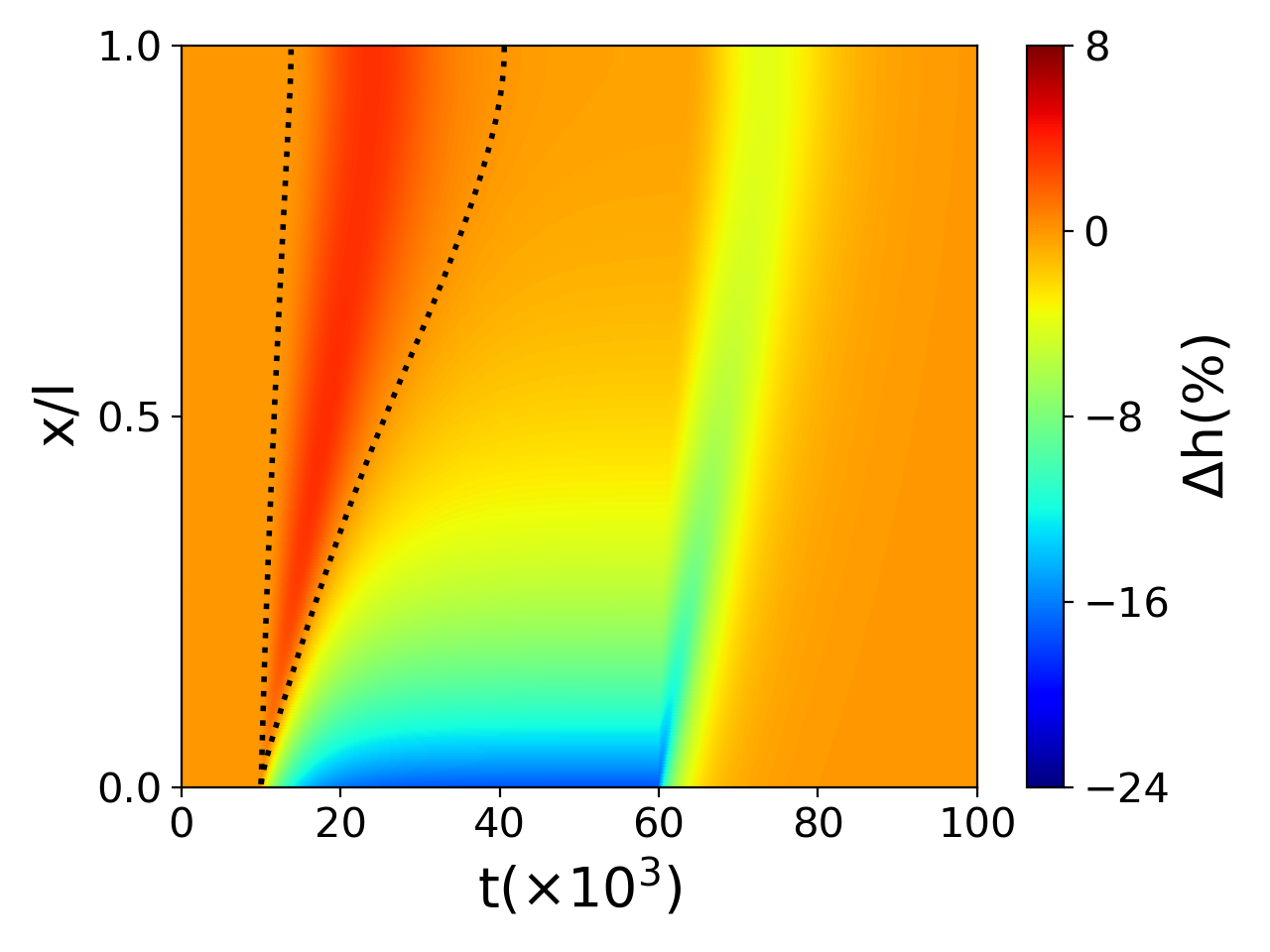}
\caption{}
\label{density_h_h20k_Fr015_S1e_4_Pe1e_2_B2e_4_mdr3_Exp1e_2.png}
\end{subfigure}
\begin{subfigure}[b]{.32\textwidth}
\includegraphics[width=\textwidth]{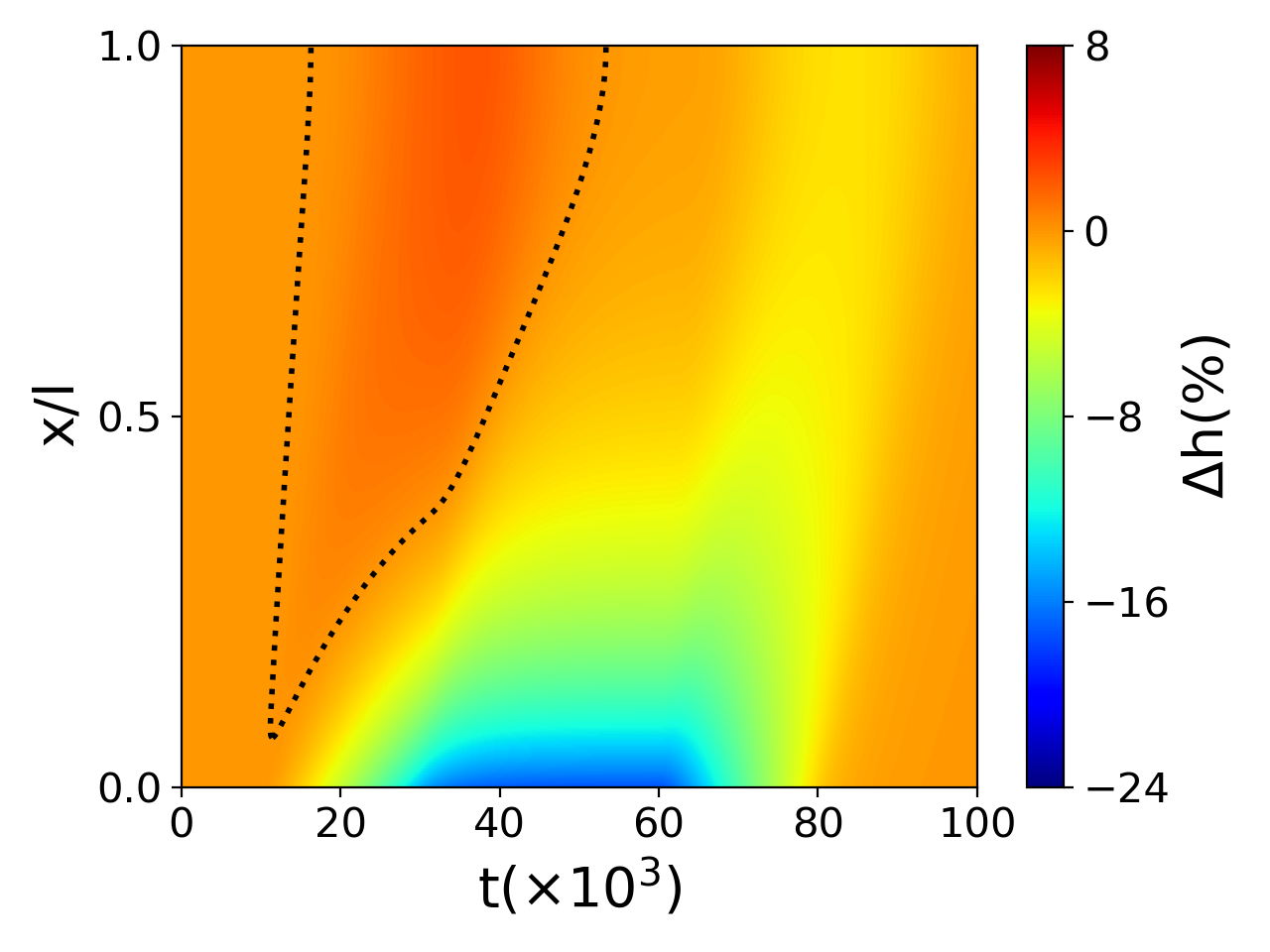}
\caption{}
\label{density_h_h20k_Fr015_S1e_4_Pe1e_2_B2e_4_mdr3_Exp1e_2_Tinj20k.png}
\end{subfigure}
\begin{subfigure}[b]{.32\textwidth}
\includegraphics[width=\textwidth]{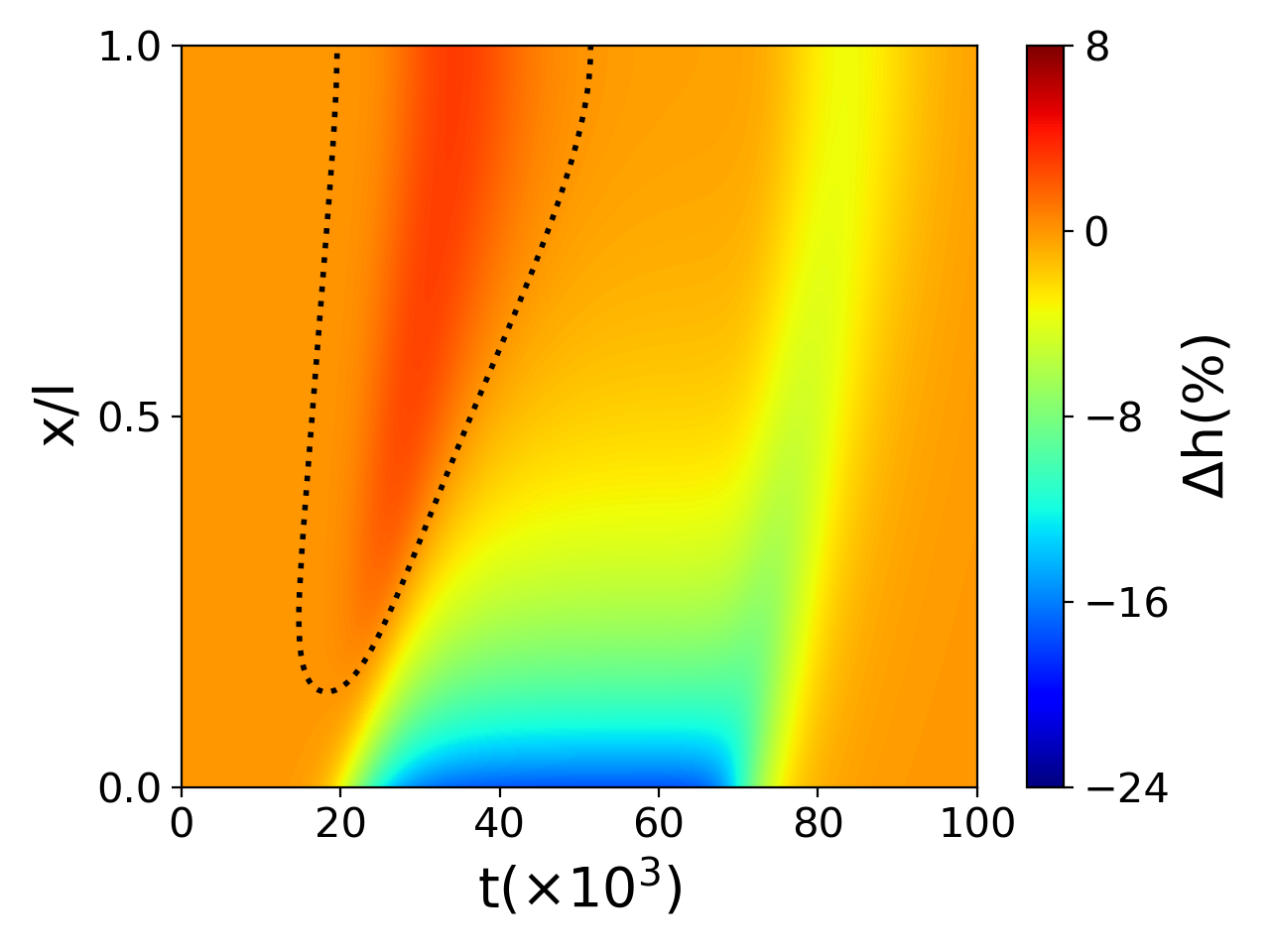}
\caption{}
\label{density_h_h20k_Fr015_S1e_4_Pe1e_2_B2e_4_mdr3_Exp1e_2_Tinj20k_tanh.png}
\end{subfigure}
\caption{ Spatiotemporal contours of change in water height for (a,d) $T_{inj}^{trans}=0$, (b,e) $T_{inj}^{trans}=20 \times 10^3 $ (linear), and (c,f) $T_{inj}^{trans}=20 \times 10^3 $ (smooth) at (a-c) $Pe \to \infty $ and $B=0$, and (d-f) $Pe = 10^{-2}$ and $B=2 \times 10^{-4}$. The values of other parameters are $Fr=0.15$, $DR_{max}=67\%$, and $l=2 \times 10^4$.}
\label{density_different_Tinj}
\end{figure}

\subsection{Mitigation of downstream height increase}
\begin{figure}[!ht]
 \centering
 \includegraphics[width=.5\textwidth]{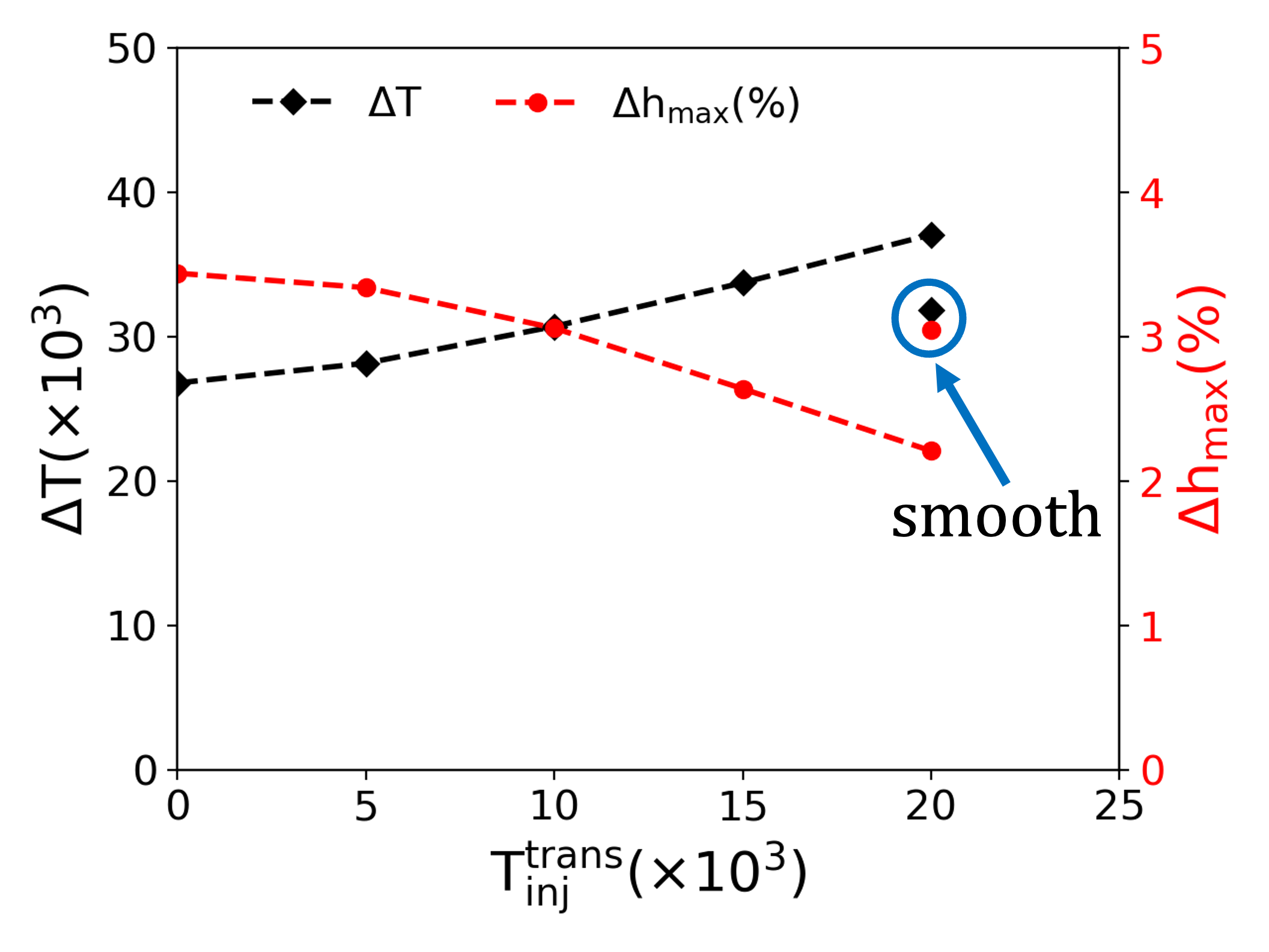}
 \caption{\label{dt_dhmax_vs_Tinj_h20k_Fr015_S1e_4_Pe1e_2_B2e_4_mdr3_Exp1e_2.png} Time duration of overshoot ($\Delta T$) and the maximum rise of water level during the overshoot at the exit of the channel for different values of $T_{inj}^{trans}$ (linear). {The data points inside the circle correspond to a smooth change in the injection concentration.} The values of other parameters are $Fr=0.15$, $Pe = 10^{-2}$, $B=2 \times 10^{-4}$, $DR_{max}=67\%$, and $l=2 \times 10^4$.}
 \end{figure}

Finally, we turn to the question of whether the undesirable \textit{increase} in water height ahead of the polymer injection front can be mitigated by changing the temporal profile of injection (Fig. \ref{cin_vs_t_Tinj20k_tnond.png}).
Spatiotemporal contours of the change in water height for sudden and gradual changes in injection concentration have been shown in Fig. \ref{density_different_Tinj}. The water height overshoot begins at the injection point for the sudden change in injection concentration (Fig. \ref{density_h_h20k_Fr015_S1e_4_Pe1e10_B0e_4_mdr3_Exp1e_2.png}), however, it initials at a place downstream of the injection point for a gradual change in injection concentration (Figs. \ref{density_h_h20k_Fr015_S1e_4_Pe1e10_B0e_4_mdr3_Exp1e_2_Tinj20k.png} and \ref{density_h_h20k_Fr015_S1e_4_Pe1e10_B0e_4_mdr3_Exp1e_2_Tinj20k_tanh.png}). {Thus, the height overshoot in the vicinity of the injection point can be completely suppressed through a gradual change in the injection concentration.} Further, the overshoot height ($\Delta h$) for the gradual change in the injection concentration is lower than the sudden injection concentration change. However, the overshoot duration for the gradual change is longer than the sudden injection concentration change. {The height overshoot also depends on the profile of the gradual change in concentration (Fig. \ref{cin_vs_t_Tinj20k_tnond.png}). The overshoot initiates further downstream of the injection point for a smooth gradual change in injection concentration (Fig. \ref{density_h_h20k_Fr015_S1e_4_Pe1e10_B0e_4_mdr3_Exp1e_2_Tinj20k_tanh.png}) compared to a linear concentration change (Fig. \ref{density_h_h20k_Fr015_S1e_4_Pe1e10_B0e_4_mdr3_Exp1e_2_Tinj20k.png}). Polymeric dispersion and degradation do not affect the qualitative nature of overshoot mitigation through the gradual change in injection concentration (Figs. \ref{density_different_Tinj}(d-f)).}
The overshoot height and duration at the exit of the channel have been quantified for different values of $T_{inj}^{trans}$ (Fig. \ref{dt_dhmax_vs_Tinj_h20k_Fr015_S1e_4_Pe1e_2_B2e_4_mdr3_Exp1e_2.png}). The time duration required to achieve the maximum injection concentration has been indicated by $T_{inj}^{trans}$ (Fig. \ref{cin_vs_t_Tinj20k_tnond.png}). The maximum height during the overshoot ($\Delta h_{max}$) decreases monotonically as the rate of the change in the injection concentration decreases (i.e., $T_{inj}^{trans}$ increases). However, the overshoot duration ($\Delta T$) increases with the increasing value of $T_{inj}^{trans}$ exhibiting a reverse trend compared to $\Delta h_{max}$. {The smoothing of time-dependent injection concentration profile close to minimum ($c_{in}=0$) and maximum ($c_{in}=1$) concentration leads to a steeper profile at the intermediate concentration (Fig. \ref{cin_vs_t_Tinj20k_tnond.png}). Therefore, the overshoot duration decreases however overshoot height goes up for a smooth change in the injection concentration compared to a linear injection concentration change (Fig. \ref{dt_dhmax_vs_Tinj_h20k_Fr015_S1e_4_Pe1e_2_B2e_4_mdr3_Exp1e_2.png}). Thus, the choice of injection profile depends on the desired objective: (i) if the desired objective is to reduce the overshoot duration ($\Delta T$) downstream of the channel then the injection profile with a sudden change in concentration is recommended (Fig. \ref{density_h_h20k_Fr015_S1e_4_Pe1e_2_B2e_4_mdr3_Exp1e_2.png}), (ii) if the desired objective is to reduce the overshoot height ($\Delta h_{max}$) downstream of the channel then the injection profile with a gradual (linear) change in concentration is desirable (Fig. \ref{density_h_h20k_Fr015_S1e_4_Pe1e_2_B2e_4_mdr3_Exp1e_2_Tinj20k.png}), and (iii) if the desired objective is to completely suppress overshoot in the vicinity of the injection location then the injection profile with a smooth gradual change in concentration is recommended (Fig. \ref{density_h_h20k_Fr015_S1e_4_Pe1e_2_B2e_4_mdr3_Exp1e_2_Tinj20k_tanh.png}).} 

\section{Conclusion}

The flow of dilute polymeric solution through an open channel has important applications in irrigation canals, sewer systems, and flood remediation. The injection of the polymer into the flow through an open channel leads to an undesirable initial overshoot of the water height before the water level goes below the base height and achieves the desired shallower steady-state depth. The origin of this overshoot lies in the increase in momentum of the fluid when the friction is decreased, which pushes the lower-momentum fluid ahead of it, leading to piling up.
{The onset of overshoot travels with the speed of the gravity wave which is much faster than the polymeric advection in the subcritical flows.} 
The water level starts retracting after the polymer's arrival and ultimately achieves a steady state. The time duration of the overshoot increases downstream of the channel. The retraction rate of water level after the arrival of polymer is immensely influenced by the local polymeric concentration. Therefore, the overshoot duration increases with the increasing value of the polymeric degradation rate. In fact, the height overshoot far downstream of the channel lasts for a longer time than the duration of the experiment for a sufficiently large polymeric degradation rate and hence the water level far downstream can remain higher than the base height throughout the experiment. Turbulent dispersion mitigates the effect of polymeric degradation and hence the duration of height overshoot decreases as the dispersion increases. The overshoot time increases as the maximum drag reduction of polymer additives increases because a larger drag reduction leads to a more severe height overshoot. {The height overshoot can lead to temporary flooding downstream of the polymeric injection point and therefore it can be detrimental for applications like flood remediation and sewer systems. However, the overshoot can be mitigated throughout the channel length and even suppressed in the vicinity of the injection location by a gradual change in the injection concentration instead of a sudden one.}

\section{Appendix}

\subsection{Numerical tool validation} \label{tool_valid}
\begin{figure}[!ht]
 \centering
 \includegraphics[width=.6\textwidth]{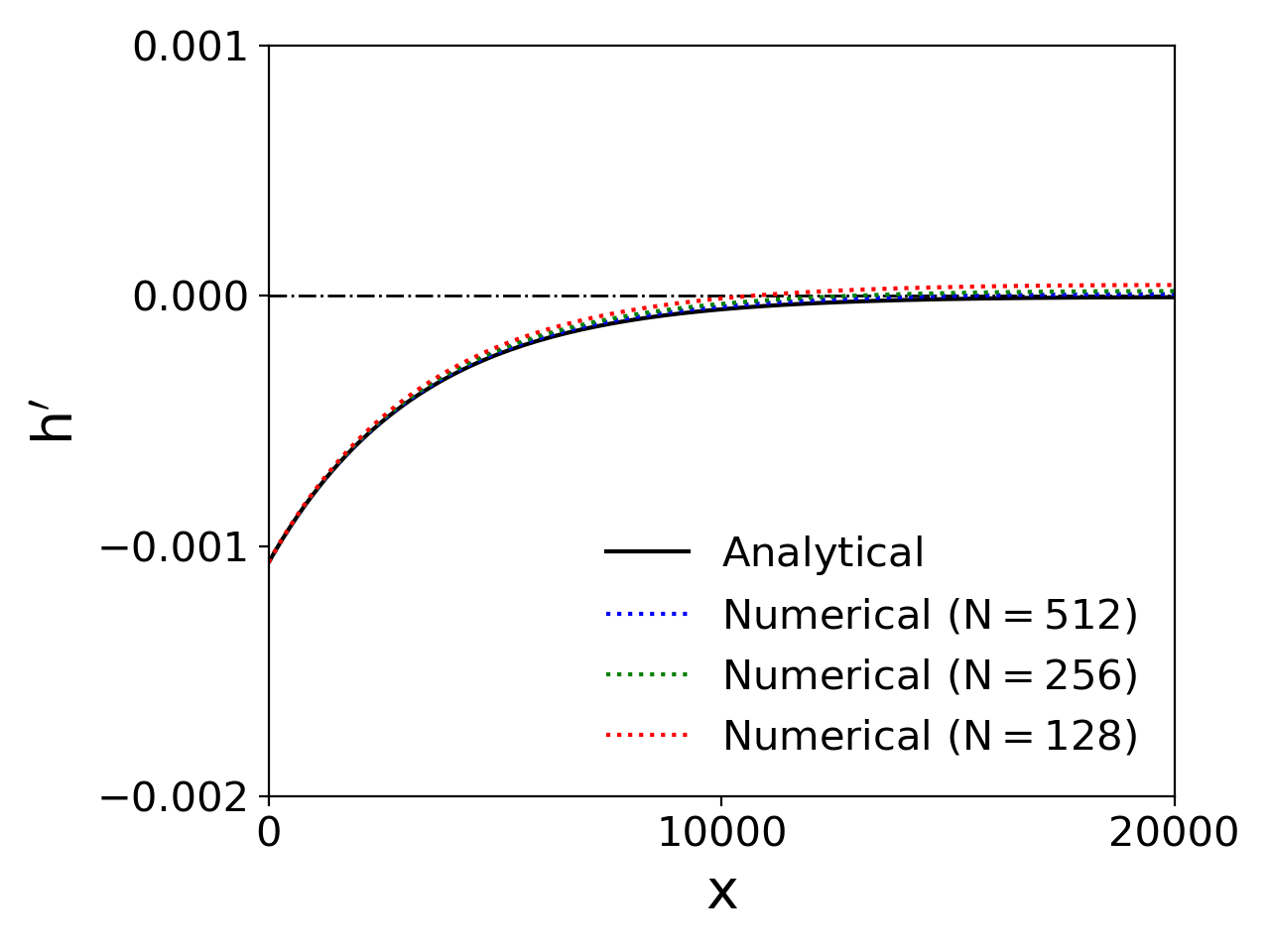}
 \caption{\label{validation_steady_height_prime_vs_x_B3e_4_N512.png} Steady state height of water level obtained from analytical and numerical methods for a tiny injection concentration of polymer ($c_{in}=0.01$). The values of other parameters are $Fr=0.15$, $Pe \to \infty$, $B=3 \times 10^{-4}$, and $DR_{max}=67\%$.}
 \end{figure}

For a small injection concentration of polymer additives, the governing equations can be linearized around the base profiles of different variables. To obtain linear governing equations, we substitute $h=h_b+h^{\prime}$, $u=u_b+u^{\prime}$, and $c=c^{\prime}$, where $h_b=1$ and $u_b=1$ are the base water height and velocity respectively, whereas the small perturbations in height, velocity, and concentration have been given as $h^{\prime}$, $u^{\prime}$, and $c^{\prime}$ respectively. The linearized equations of the conservation of mass, momentum, and concentration can be given as: 

\begin{equation}\label{com_lin}
\frac{\partial h^{\prime}} {\partial t}+\frac{\partial h^{\prime}} {\partial x}+\frac{\partial u^{\prime}} {\partial x}=0,
\end{equation}

\begin{equation}\label{colm_lin}
\frac{\partial u^{\prime}} {\partial t}+\frac{\partial u^{\prime}} {\partial x}+\frac{1} { Fr^2} \frac{\partial h^{\prime}} {\partial x}+\frac{1} { Fr^2} \frac{n^2} { R_b^{4/3}} \left \{2u^{\prime} -\frac{4}{3}h^{\prime}- (DR_{max})c^{\prime} \right\}=0,
\end{equation}
and
\begin{equation}\label{coc_lin}
\frac{\partial c^{\prime}} {\partial t}+\frac{\partial c^{\prime}} {\partial x}-\frac{1} {Pe} \frac{\partial^2 c^{\prime} } {\partial x^2}+B c^{\prime}=0,
\end{equation}
where $R_b$ is the base hydraulic radius of the channel. The linearized governing equations can be solved analytically for steady flow. The water level heights at steady state obtained from numerical method with different mesh resolutions have been shown along with the analytical solution in Fig. \ref{validation_steady_height_prime_vs_x_B3e_4_N512.png}. The numerical simulations throughout the study have been performed with $N=512$ as this resolution is sufficient for the mesh-independent results.   

\section*{Declaration of competing interest}
There is no competing interest to declare. 

\section*{Acknowledgments}
The authors acknowledge support from ONR N00014-18-1-2865 (Vannevar Bush Faculty Fellowship), and helpful discussions with Nimish Pujara.

\bibliography{apssamp}
\end{document}